\documentclass[reqno,11pt]{amsart}
\usepackage[linktocpage]{hyperref}
\usepackage[utf8]{inputenc}
\usepackage{graphicx}
\usepackage{slashed}
\usepackage{amscd}
\usepackage{amssymb}
\usepackage{esint}
\usepackage{enumitem}
\usepackage{comment}
\usepackage{amsmath}
\usepackage{mathtools}
\usepackage{dsfont}
\usepackage[mathscr]{eucal}
\usepackage{epsfig}
\usepackage{pstricks}
\usepackage{pst-all}
\usepackage{soul}
\usepackage{dirtytalk}
\usepackage{marginnote}
\usepackage{cleveref}
\usepackage[gen]{eurosym}
\usepackage[
  backend=biber,
  style=trad-abbrv,
  sorting=none
]{biblatex}
\usepackage{tikz-cd}
 \usetikzlibrary{calc}
\usepackage{acro}
\usepackage{float}

\DeclareAcronym{cfs}{
  short = CFS,
  long = causal fermion systems
}
\DeclareAcronym{gr}{
  short = GR,
  long = general relativity
}
\DeclareAcronym{qft}{
  short = QFT,
  long = quantum field theory
}
\DeclareAcronym{qt}{
  short = QT,
  long = quantum theory
}
\DeclareAcronym{qrf}{
  short = QRF,
  long = quantum reference frame,
  short-plural = s,
  long-plural = s
}

\numberwithin{equation}{section}
\allowdisplaybreaks[1]
\definecolor{labelkey}{gray}{.65}

\title[Observables and Symmetries in CFS]{Observables and Symmetries in Causal Fermion System}

\author[P.\ Fischer]{Patrick Fischer$^{\dagger}$}
\email{patrick.fischer@ur.de}

\author[C.F.\ Paganini]{Claudio F. Paganini$^{\dagger}$}
\email{claudio.paganini@mathematik.uni-regensburg.de}
\address{$^{\dagger}$ Fakultät für Mathematik \\ Universität Regensburg \\ D-93040 Regensburg \\ Germany}

\author[S.\ Yadav]{Sahil Yadav$^{*}$}
\email{sahilshubham2003@outlook.com}
\address{$^{*}$ Department of Physics \\ Stockholm University \\ Roslagstullsbacken 21 \\ Stockholm \\ Sweden}

\newtheorem{Def}{Definition}[section]
\newtheorem{Thm}[Def]{Theorem}
\newtheorem{Prp}[Def]{Proposition}
\newtheorem{Lemma}[Def]{Lemma}
\newtheorem{Remark}[Def]{Remark}
\newtheorem{Corollary}[Def]{Corollary}

\newtheorem{example}{Example}

\renewcommand{\H}{\mathscr{H}}
\newcommand{\F}{\mathscr{F}}
\renewcommand{\L}{\mathcal{L}}
\renewcommand{\l}{\lambda}
\newcommand{\N}{\mathbb{N}}
\newcommand{\C}{\mathbb{C}}
\renewcommand{\S}{\mathcal{S}}
\renewcommand{\O}{\mathcal{O}}
\newcommand{\bra}{\langle}
\newcommand{\ket}{\rangle}
\newcommand{\R}{\mathbb{R}}
\newcommand{\M}{\mathcal{M}}

\newcommand{\tr}{\operatorname{tr}}
\renewcommand{\and}{\quad\text{and}\quad}

\renewcommand{\tr}{\operatorname{tr}}
\renewcommand{\Tr}{\operatorname{Tr}}
\renewcommand{\and}{\quad\text{and}\quad}

\newcommand{\G}{\mathscr{G}}

\newcommand\mpar[1]{}
\begin{document}

\begin{abstract}
    We investigate the interplay of observables and symmetries in the framework of \ac{cfs}. We show that position observables transform covariantly under unitary symmetries and for static \ac{cfs} we derive their Heisenberg-type evolution. To give these results a geometric interpretation, we establish conditions under which spacetime isometries induce symmetries of \ac{cfs} constructed from solutions of the Dirac equation. The crucial link is provided by the Dirac equation itself, which relates the infinitesimal generators of the geometric and \ac{cfs} symmetries. In static spacetimes, this allows us to identify the generator of the induced unitary symmetry with the spatial Dirac Hamiltonian.
\end{abstract}

\maketitle


\acreset{cfs}

\tableofcontents

\section{Introduction} \label{sec:introduction}
The framework of \ac{cfs} provides a unified description of spacetime and quantum matter in terms of a measure on a space of linear operators \cite{CFS_Book, Unified_Physical_Theory, CFS_Physical_Mathematical_Idea}. The operators encode the local correlations of quantum states, and spacetime is identified with the support of the measure. The dynamics are governed by the causal action principle. In the continuum limit, the equations of general relativity and the gauge field interactions of the Standard Model emerge from its Euler–Lagrange equations \cite{Gravity, Continuum_Limit_CFS}. Recently, the continuum limit analysis has been extended to general globally hyperbolic spacetimes \cite{Continuum_FF}. Moreover, quantum field theory has been recovered in a suitable limiting regime \cite{QFT, Fock_Space}. The framework has also given rise to novel phenomenological proposals, including a mechanism for baryogenesis and effective collapse models \cite{Baryon, Collapse, Collapse-2}.
Taken together, these developments establish \ac{cfs} as a leading candidate for a unified theory of fundamental physics.

Symmetries play a fundamental role in modern theoretical physics. In \cite{Preserver_Paper}, it was shown that, for finite-dimensional Hilbert spaces, continuous maps preserving the causal action principle are induced by unitary or antiunitary transformations of the underlying Hilbert space. Correspondingly, symmetries of a given \ac{cfs} are realized by such transformations that leave its measure invariant \cite{Positive_Mass, Noether-like, Langer_Master_Thesis}. Symmetries have also been investigated in the context of causal variational principles, where Noether-like theorems give rise to conservation laws expressed through surface layer integrals \cite{Noether-like, Hamiltonian_Structure, Surface_Layer_Integral}. More recently, position observables have been introduced, associating operators with spacetime regions \cite{PC}. This development also opens connections to relational observables and quantum reference frames \cite{QRF_CG, Leon_QRF, Leon_Relative_Time, QRF-1, QRF-2, QRF-3}.

While position observables in \ac{cfs} can be associated with arbitrary spacetime regions, there is no a priori prescription for relating observables at different times. To make dynamical predictions, we therefore need to identify how these observables evolve. A one-parameter group of symmetries provides precisely such a relation through its action on spacetime. We show that the invariance of the measure under unitary conjugation implies the covariance of position observables. For a strongly continuous one-parameter group of symmetries, this covariance leads to a Heisenberg--type evolution equation generated by the corresponding self-adjoint operator. In static \ac{cfs}, where the symmetry acts by time translations, this gives the familiar form of temporal evolution for position observables. Effective collapse models provide an interesting contrast, as stochastic perturbations can break the symmetry in individual realizations, even when it is retained at the ensemble level \cite{Collapse, Collapse-2}. Motivated by the role of commutation relations in these models, we also investigate sufficient conditions under which position observables commute with the symmetry generator.

To investigate symmetries in \ac{cfs} constructed from Lorentzian spacetimes, we draw on the extensive mathematical literature on the Dirac equation. Its theory on curved Lorentzian spacetimes is well developed, including the well-posedness of the Cauchy problem on globally hyperbolic spacetimes, finite speed of propagation of solutions \cite[Theorem 4.5.2]{CFS_Book}, and the construction of propagators \cite{Dirac_Operator-2, Cauchy_Problem, Boundary_Value_Problem,Propagator,Dirac_on_Manifold}. The spectral theory of Dirac operators and their Hamiltonian formulations, particularly on stationary and static spacetimes, have likewise been investigated extensively \cite{Islam2023Gutzwiller, Self-adjoint_Dirac_Hamiltonian,Dirac_Hamiltonian_Unique}. The action of spacetime isometries on spinor fields and the associated spinorial Lie derivatives are also well understood \cite{Compatible_Lift, Godina_and_Matteucci, Dirac_Operator_on_Riemann}. These constructions rely on established functional-analytic tools, including strongly continuous unitary groups, self-adjoint generators, and invariant-core arguments \cite{ST, Engel_Nagel, ReedSimon1972, Self_adjointness}. Particularly important for our purposes is the conserved Dirac current, which gives rise to a positive-definite scalar product on solutions that is independent of the choice of Cauchy hypersurface \cite{Dirac_Operator-2, Thaller1992, Dirac_on_Manifold},\cite[Section~4]{CFS_Book}. Together with the well-posedness of the Cauchy problem, this provides canonical unitary identifications between the Hilbert space of Dirac solutions and the Hilbert spaces of initial data on different Cauchy hypersurfaces. In this work, we carefully distinguish these Hilbert spaces and systematically identify the different maps between them, including Cauchy evolution, geometric transport under spacetime isometries, and the induced unitary transformations. The induced unitary action of spacetime symmetries on Dirac solutions has already been investigated in the \ac{cfs} context by Finster and Reintjes \cite{SigSymm}, who studied its implications for the fermionic signature operator and its spectral subspaces.

Building on these mathematical foundations, we establish sufficient conditions under which spacetime isometries induce symmetries of \ac{cfs}. Starting from a globally hyperbolic Lorentzian spin spacetime, we show that Dirac-compatible spin isometries induce unitary transformations on the Hilbert space of Dirac solutions. Given an invariant Hilbert subspace on which the local correlation map is well-defined, we prove that this map intertwines the spacetime isometry $\Phi$ with the induced unitary transformation $U$, according to
\[
    F(\Phi(x))=UF(x)U^{-1}.
\]
Together with the invariance of the metric volume measure under spacetime isometries, this implies that $U$ preserves the measure of the resulting \ac{cfs} and therefore defines a symmetry in the sense of \ac{cfs}. For complete Killing flows, we obtain strongly continuous one-parameter groups of such symmetries. Crucially, the Dirac equation provides the connection between the infinitesimal generators of the geometric and \ac{cfs} symmetries. In static spacetimes, it relates the spinorial Lie derivative along the Killing vector field to the spatial Dirac Hamiltonian, allowing us to identify the generator of the induced unitary symmetry with the corresponding Hamiltonian on the invariant subspace. Different choices of spin lift can shift this generator by a constant multiple of the identity, but leave the induced \ac{cfs} symmetry unchanged. Combined with the covariance of position observables, these results establish a direct connection between spacetime geometry, Dirac dynamics, and the evolution of observables in \ac{cfs}.

Our results also clarify how internal symmetries are represented in \ac{cfs}. In spin geometry, internal symmetries act on the fibres of the spinor bundle while leaving the base points of the spacetime manifold invariant. We consider such transformations that preserve Clifford multiplication and the spin scalar product. If they are also compatible with the Dirac equation and preserve the chosen Hilbert subspace, the equivariance of the local correlation map implies
\[
    [U,F(x)]=0 \qquad \text{for all }x\in M.
\]
Thus, internal symmetries leave every local correlation operator invariant and consequently fix the spacetime of the resulting \ac{cfs} pointwise. This distinguishes their representation in \ac{cfs} from that of spacetime isometries, which generally act nontrivially on the local correlation operators. For an irreducible Dirac representation, the internal symmetries satisfying these conditions are restricted to global $U(1)$ phase transformations, whereas multiple equivalent Dirac sectors allow larger internal symmetry groups. Of particular interest in this context are previous investigations of internal symmetries associated with octonionic structures \cite{Octonian}. Our results show that such symmetry structures can be transferred to the corresponding \ac{cfs}, provided they are represented by compatible transformations of the Dirac sectors and the regularization preserves these symmetries. In particular, choosing identical regularizations for equivalent Dirac sectors ensures invariance under internal transformations that mix these sectors. By contrast, local Lorentz transformations merely change the choice of geometric frames and therefore do not appear as distinct transformations of the resulting \ac{cfs}.

\subsubsection*{\textbf{Organization}}

The paper is organized as follows. In Section~\ref{subsec:preliminaries_cfs} we review the basic definitions of \ac{cfs} and introduce the notion of unitary symmetry. In Section~\ref{sec:observable_&_evolution} we establish the covariance of position observables, derive their Heisenberg-type evolution, and investigate commutation relations motivated by effective collapse models. Section~\ref{sec:symmetry_mapping} begins with the Dirac Cauchy problem and a systematic treatment of the Hilbert spaces and unitary maps involved in spacetime symmetries. We then establish the correspondence between spacetime isometries and \ac{cfs} symmetries through the local correlation map, discuss the representation of internal symmetries, and identify the symmetry generator with the spatial Dirac Hamiltonian in the static setting. We illustrate the construction with three examples involving finite- and infinite-dimensional Hilbert spaces. Finally, Section~\ref{sec:conclusion_&_outlook} summarizes our results and discusses directions for future research.

\subsubsection*{\textbf{Acknowledgment}} We would like to thank Felix Finster, Simone Murro and Antonino Marciano for helpful discussions and feedback on earlier versions of the manuscript. S.Yadav would also like to thank Ingemar Bengtsson for many helpful discussions throughout this project. Discussions with large language models were used throughout the preparation of this manuscript. All scientific conclusions, proofs, and final text were independently verified and authored by the researchers, who assume complete responsibility for the entirety of the work.

\section{Preliminaries on Causal Fermion Systems}\label{subsec:preliminaries_cfs}

We begin with the formal definition of \ac{cfs} according to \cite[Definition~5.4.1]{CFS_Book},

\begin{Def}[Causal Fermion System] \label{def:cfs_definition}
    Let~$\H$ be a separable complex Hilbert space together with a scalar product~$\bra \cdot , \cdot \ket_\H$, and denote by $\mathrm{L}(\H)$ the set of linear operators on~$\H$.  Moreover, given a parameter~$n \in \N$ $($the so-called {\bf{\say{spin dimension}}}$)$, we let $\F_{n} \subset \mathrm{L}(\H)$ be the set of all self-adjoint operators on~$\H$ of finite rank, which $($counting multiplicities$)$ have at most $n$ positive and at most $n$ negative eigenvalues. Furthermore, let $\rho$ be a positive measure on $\F_{n}$ $($defined on a $\sigma$-algebra of subsets of $\F_{n}$$)$, the so-called {\bf{measure}}. We refer to $(\H, \F_{n}, \rho)$ as a {\bf{causal fermion system}}.
\end{Def}

For any $x,y \in \F_{n}$, the operator product $xy$ has rank at most $2n$ with the eigenvalues (counting algebraic multiplicities), $\l^{xy}_1, \ldots, \l^{xy}_{2n} \in \C$. The causal action is given as a function of the eigenvalues as,
\begin{align} \label{eq:causal_action}
	\S(\rho) 
    =
    \iint_{\F_{n} \times \F_{n}} \L(x,y)\: d\rho(x)\, d\rho(y) \:. 
\end{align}
where the causal Lagrangian is,
\begin{equation} \label{eq:causal_lagrangian}
    \L(x,y) 
    =
    \frac{1}{4n}\sum_{i,j=1}^{2n}
    \left( 
    |\lambda^{xy}_{i}|-
    |\lambda^{xy}_{j}|
    \right)^{2}
\end{equation}
The {\emph{causal action principle}} is to minimize $\S$ by varying the measure $\rho$ under the following constraints:
\begin{align} 
    \text{\emph{volume constraint:}} 
    && 
    \rho(\F_{n}) 
    =
    1 
    \quad\;\; \label{eq:volume_constraint}
    \\ 
    \text{\emph{trace constraint:}} 
    && 
    \Tr(\rho) 
    :=
    \int_{\F_{n}} \tr(x)\: d\rho(x) 
    =
    1
    &
    \\ 
    \text{\emph{boundedness constraint:}} 
    && 
    \mathcal{T}(\rho) 
    := 
    \iint_{\F_{n} \times \F_{n}} 
    \left(
    \sum^{2n}_{i=1}|\lambda_i^{xy}|
    \right)^{2}
    \: d\rho(x)\, d\rho(y) 
    & 
    <
    \infty \:, 
\end{align}
Furthermore, the notion of causality in \ac{cfs} is defined according to \cite[Definition 5.6.1]{CFS_Book} as,

\begin{Def}[Causal Structure]\label{def:causal_structure}
    For the product $xy$, points $x$ and $y$ are called {\bf{space--like}} separated if all the $\lambda^{xy}_{j}$ have the same absolute value. They are said to be {\bf{time--like}} separated if the $\lambda^{xy}_{j}$ are all real and do not all have the same absolute value. In all other cases, the points $x$ and $y$ are said to be {\bf{light--like}} separated.
\end{Def}

We can see that the notion of causality is compatible with the causal Lagrangian~\eqref{eq:causal_lagrangian}. Suppose, $x,y\in \F_{n}$ are space-like separated then the eigenvalues $\lambda^{xy}_{j}$ have the same absolute values. As a consequence, the causal Lagrangian~\eqref{eq:causal_lagrangian} vanishes. Thus, the space-like separated points do not contribute in the causal action~\eqref{eq:causal_action}.

Having introduced the causal structure and the causal action principle, we now turn to the geometric objects that encode the physical content of a \ac{cfs}. In particular, we introduce the spin spaces and the associated physical wave functions, which provide the link between the Hilbert space and the emergent spacetime. 

For every $x \in \F_{n}$ we define the spin space $S_{x}$ as the image of the operator $x$, i.e. $S_{x} := x(\H)$ it is a subspace of $\H$ of dimension at most $2n$. Moreover, let $\pi_{x} : \; \H \rightarrow S_{x}$ be the orthogonal projection in $\H$ on the subspace $S_{x} \subset \H$. Then, for a vector $u \in \H$, one introduces the \emph{physical wave function} $\psi^{u}$ as,
\begin{align*}
    \psi^{u}:\;M\rightarrow \H,
    \quad\quad
    \psi^{u}(x)
    :=
    \pi_{x}u \in S_{x}
\end{align*}
where, $M := \operatorname{supp} \rho \subset \F_{n}$, which is interpreted as the \emph{spacetime}.

Symmetries are fundamental to modern physics, governing both the dynamics of physical systems and the structure of their observables. We conclude these preliminaries by introducing the corresponding notion for \ac{cfs}.

\begin{Def}[Symmetry of a \ac{cfs}] \label{def:symm}
    Let $(\H,\F_n,\rho)$ be a \ac{cfs}. A unitary operator $U$ on $\H$ is called \textbf{a symmetry of the \ac{cfs}} if
    \[
        \rho(U\Omega U^{-1})
        =
        \rho(\Omega)
    \]
    for every Borel set $\Omega \subset \F_n$, where
    \[
        U \Omega U^{-1}
        :=
        \{UxU^{-1} \mid x \in \Omega\}.
    \]
\end{Def}

Finally, in \cite[Definition~6.2]{Static_CFS} an important special class of symmetric \ac{cfs} was introduced.

\begin{Def}[Static \ac{cfs}] \label{def:static_CFS}
    Let $\{U_{t}\}_{t\in\R}$ be a strongly continuous one--parameter group of unitary transformations on the Hilbert space $\H$. The causal fermion system $(\H,\F_n,\rho)$ is static with respect to $\{U_{t}\}_{t \in \R}$ if it has the following properties:
    \begin{enumerate}
        \item Space-time $M := \operatorname{supp} \rho \subset \F_n$ is a topological product,
        \[
            M 
            =
            \R \times \mathcal{N} 
            \quad\text{or}\quad 
            M 
            =
            S^{1} \times \mathcal{N}.
        \]
        We write a space-time point $x \in M$ as $x = (t,\mathbf{x})$ with $t \in \R$ and $\mathbf{x} \in \mathcal{N}$, where in the case $M= S^{1} \times \mathcal{N}$ we identify the circle with the unit interval via 
        \[
            S^{1} \simeq \R \mod \mathbb{Z}.
        \]
    \item The one-parameter group $\{U_{t}\}_{t \in \R}$ leaves the measure $\rho$ invariant, i.e. 
    \[
        \rho(U_{t} \Omega U^{-1}_{t}) 
        = 
        \rho(\Omega) 
        \qquad
        \text{for all $\rho$-measurable sets $\Omega\subset \F_n$}.
    \]
    Moreover,
    \[
        U_{t'} (t,\mathbf{x}) U^{-1}_{t'} 
        =
        (t + t', \mathbf{x})
    \]
    (where in the case $M = S^{1} \times \mathcal{N}$ the sum $t + t'$ is taken modulo $\mathbb{Z}$).
    \end{enumerate}
\end{Def}

In the next section, we'll introduce the position observable and study its evolution for symmetric \ac{cfs}.

\section{Observables and Time Evolution} \label{sec:observable_&_evolution}

In any physical theory, observables play a central role. Within the framework of \ac{cfs}, the notion of a position observable is introduced in \cite[Definition~3.5]{PC} as following,

\begin{Def}[Position Observable] \label{def:observables}
    Let $(\H, \F_{n}, \rho)$ be a \ac{cfs} and let $\mathcal{A} \subseteq \F_{n}$. The \textbf{observable} associated with the region $\mathcal{A}$ is defined as the operator
    \begin{align}\label{obser}
        \O(\mathcal{A}) 
        := 
        \int_{\mathcal{A}} \pi_x \, d\rho(x),
    \end{align}
    where $\pi_x$ denotes the orthogonal projection onto the image of $x$.
\end{Def}

The position observable satisfies certain fundamental properties, summarized in the following proposition \cite[Proposition 3.6]{PC}.

\begin{Prp} \label{prp:observables_prop}
    Let $\mathcal{A} \subseteq \F_{n}$. Then the observable $\O(\mathcal{A})$ satisfies the following property,
    \begin{enumerate}[label=(\arabic*)]
        \item $M \cap \mathcal{A} = \varnothing \implies \O(\mathcal{A}) = 0$.
        \item $\O(\mathcal{A})=0\implies\rho\big((M\cap \mathcal{A})\setminus\{0\}\big)=0$.
    \end{enumerate}
\end{Prp}

\subsection{Time Evolution of Observable in Static \ac{cfs}}

Having introduced observables, a next natural step is the study of their evolution. In this section, we focus on the time evolution of the position observable. For this, it is necessary to have the notion of symmetry within the framework of \ac{cfs} defined as Definition~\ref{def:symm}.

Additionally, this definition extends naturally to a one-parameter family of unitary transformations $\{U_{t}\}_{t \in \mathbb{R}}$.

The position observable $\O$ is defined as an integral over projection operators $\pi_x$, Definition~\ref{def:observables}. To analyse its unitary evolution, we first establish how projection operators transform under unitary evolution.

\begin{Lemma} \label{projection}
    Let $\H$ be a Hilbert space and $x$ a self-adjoint operator on $\H$. Let $\pi_x$ be the projection onto the image of $x$. Let $U$ be a unitary transformation on $\H$. Then the following holds
    \begin{equation}
        U \pi_x U^{-1} 
        =
        \pi_ {U x U^{-1}}.
    \end{equation}
\end{Lemma}

\begin{proof}
    Since $x$ is a self-adjoint operator on the Hilbert space $\H$, its image $\operatorname{im}(x)$ is the closed subspace spanned by eigenvectors with non-zero eigenvalues. The projection $\pi_x$ projects onto $\operatorname{im}(x)$. Define $x' = U x U^{-1}$, where $U$ is unitary ($U^{-1} = U^\dagger$), so $\pi_{x'}$ projects onto $\operatorname{im}(x')$.
    
    To show $\operatorname{im}(x') = U \operatorname{im}(x)$, consider $\phi = x \psi \in \operatorname{im}(x)$. Then $x' (U \psi) = U x U^{-1} (U \psi) = U x \psi = U \phi$, so $U \phi \in \operatorname{im}(x')$, implying $U \operatorname{im}(x) \subseteq \operatorname{im}(x')$. Conversely, for $\eta = x' \xi \in \operatorname{im}(x')$, let $\psi = U^{-1} \xi$. Then $\eta = x' \xi = U x U^{-1} \xi = U x \psi$, so $\eta \in U \operatorname{im}(x)$. Thus, $\operatorname{im}(x') = U \operatorname{im}(x)$.

    For $\psi = U \phi$ with $\phi \in \operatorname{im}(x)$, we have $(U \pi_x U^{-1}) \psi = U \pi_x \phi = U \phi = \psi$, so $U \pi_x U^{-1}$ is the identity on $\operatorname{im}(x')$. Thus, $U \pi_x U^{-1}$ projects onto $\operatorname{im}(x')$, as does $\pi_{x'}$. Since both are projections onto the same subspace, $U \pi_x U^{-1} = \pi_{x'} = \pi_{U x U^{-1}}$.
\end{proof}

To formulate the covariance relation for position observables, we first recall two measure-theoretic notions.
A measure \(\rho\) on \(\F_n\) is called $\sigma$-finite if there exists a countable family of Borel sets \(A_k\subset\F_n\) such that
\[
    \F_n
    =
    \bigcup_{k=1}^{\infty}A_k,
    \qquad
    \rho(A_k)<\infty
    \quad\text{for all }k.
\]
The measure is called locally finite if every \(x\in\F_n\) has an open neighbourhood \(V_x\) satisfying
\[
    \rho(V_x)<\infty.
\]
Since \(\F_n\), equipped with the operator-norm topology, is second-countable, local finiteness implies $\sigma$-finiteness. The converse does not hold in general.

\begin{Thm} \label{thm:unitary_observable}
Let \((\H,\F_n,\rho)\) be a \ac{cfs} with $\sigma$-finite measure \(\rho\), and let \(U\) be a unitary operator on \(\H\). Then \(U\) is a symmetry of the \ac{cfs} if and only if
\[
    \mathcal O(UAU^{-1})
    =
    U\mathcal O(A)U^{-1} 
    =
    \O(\mathcal{A}_{U})
\]
for every Borel set \(A\subset\F_n\) with \(\rho(A)<\infty\) and where $\mathcal{A}_{U}:= U\mathcal{A}U^{-1} = \{ U x U^{-1}| x\in \mathcal{A}\}$.
\end{Thm}

\begin{proof}
    We first show that if \(U\) is a symmetry of the \ac{cfs}, then the position observables satisfy the covariance relation.
   The observable associated with a spacetime region $\mathcal{A} \subset \F_{n}$ is defined by Definition~\ref{def:observables}. For the transformed region $U \mathcal{A} U^{-1}$, the corresponding observable is:
    \[
        \O(U \mathcal{A} U^{-1}) 
        =
        \int_{U \mathcal{A} U^{-1}} \pi_x \, d\rho(x).
    \]
    Perform a change of variables in the integral, setting $y = U x U^{-1}$, so that $x = U^{-1} y U$ and $U \mathcal{A} U^{-1} = \{ y = U x U^{-1} \mid x \in \mathcal{A} \}$. Thus, the integral becomes:
    \begin{align*}
        \O(U \mathcal{A} U^{-1}) 
        =
        \int_{\mathcal{A}} \pi_{U x U^{-1}} \, d\rho(U x U^{-1})
        \overset{\ref{def:symm}}{\Longrightarrow}
        \O(U \mathcal{A} U^{-1}) 
        =
        \int_{\mathcal{A}} \pi_{U x U^{-1}} \, d\rho(x).
    \end{align*}
    By Lemma~\ref{projection}, the operator-valued function satisfies $\pi_{U x U^{-1}} = U \pi_x U^{-1}$. Substituting this, we obtain:
    \[
        \O(U \mathcal{A} U^{-1}) 
        =
        \int_{\mathcal{A}} U \pi_x U^{-1} \, d\rho(x).
    \]
    Since \(U\) and \(U^{-1}\) are bounded and independent of the integration variable, they can be taken outside the integral
    \begin{equation}
        \O(U \mathcal{A} U^{-1}) 
        =
        U \left( \int_{\mathcal{A}} \pi_x \, d\rho(x) \right) U^{-1} 
        =
        U \O(\mathcal{A}) U^{-1}.
    \end{equation}
    Thus, the desired equality holds, $U \O(\mathcal{A}) U^{-1} = \O(U \mathcal{A} U^{-1})$.

    Conversely, suppose that
    \[
        \mathcal O(UAU^{-1})
        =
        U\mathcal O(A)U^{-1}
    \]
    holds for every Borel set \(A\subset\F_n\) with \(\rho(A)<\infty\). 
    For \(r=1,\ldots,2n\), consider the Borel subsets
    \[
        \F_n^{(r)}
        =
        \{x\in\F_n:\operatorname{rank}(x)=r\}.
    \]
    Let \(A\subset\F_n^{(r)}\) be a Borel set with \(\rho(A)<\infty\). Since unitary conjugation preserves rank,
    \[
        UAU^{-1}
        \subset
        \F_n^{(r)}.
    \]
    Moreover, for every \(x\in\F_n^{(r)}\), the orthogonal projection \(\pi_x\) has rank \(r\), and therefore
    \[
        \operatorname{tr}(\pi_x)
        =
        r.
    \]
    Using the definition of the position observable, the covariance relation, and the unitary invariance of the trace, we obtain
    \[
    \begin{aligned}
        r\,\rho(A)
        =
        \int_A\operatorname{tr}(\pi_x)\,d\rho(x)
        =
        \operatorname{tr}\mathcal O(A)
        =
        \operatorname{tr}\bigl(U\mathcal O(A)U^{-1}\bigr) 
        =
        \operatorname{tr}\mathcal O(UAU^{-1})
        =
        r\,\rho(UAU^{-1})\,.
    \end{aligned}
    \]
    The interchange of trace and integration is justified by positivity, while the covariance relation ensures that the transformed observable is trace-class.
    Since \(r>0\), it follows that
    \[
        \rho(UAU^{-1})
        =
        \rho(A).
    \]
    Thus, the measure is invariant on every finite-measure Borel subset of each positive-rank stratum. Since \(\rho\) is $\sigma$-finite, there exists a countable family of Borel sets \(A_k\) of finite measure covering \(\F_n\). By decomposing an arbitrary Borel set into disjoint subsets of the \(A_k\) and the rank strata, and applying countable additivity, the preceding equality extends to all Borel subsets of positive rank.
    Finally, the rank-zero stratum consists only of the zero operator, which is fixed under unitary conjugation:
    \[
        U0U^{-1}
        =
        0.
    \]
    Consequently,
    \[
        \rho(UAU^{-1})
        =
        \rho(A)
    \]
    for every Borel set \(A\subset\F_n\).
    By Definition \ref{def:symm}, \(U\) is therefore a symmetry of the \ac{cfs}.
\end{proof}

Before addressing the time evolution explicitly, we recall Stone’s theorem \cite[Theorem~10.15]{ST}.

\begin{Thm}[Stone's Theorem] \label{thm:Stone}
    Suppose $U_{t}$ is a strongly continuous one--parameter unitary group on $\H$. Then the infinitesimal generator $A$ of $U_{t}$ is densely defined and self-adjoint, and $U_{t} = e^{itA}$ for all $t\in \R$.
\end{Thm}

As a direct consequence, we obtain the Heisenberg equation for the position observable.

\begin{Corollary}\label{Col:Heisenberg}
    Let \((\H,\F_n,\rho)\) be a CFS with $\sigma$-finite measure \(\rho\), and let \((U_t)_{t\in\mathbb R}\) be a strongly continuous one-parameter unitary group with self-adjoint generator \(H\), such that \(U_t=e^{itH}\). For a Borel set \(A\subset\F_n\), define \(A_t:=U_tAU_t^{-1}\).
    
    Then \((U_t)_{t\in\mathbb R}\) is a one-parameter group of CFS symmetries if and only if, for every Borel set \(A\subset\F_n\) with \(\rho(A)<\infty\), the position observables \(\mathcal O(A_t)\) are well-defined as bounded operators for all \(t\in\mathbb R\), depend strongly continuously on \(t\), and satisfy the Heisenberg equation
    \begin{equation} \label{eq:Heisenberg}
        \frac{d}{dt}\O(\mathcal{A}_{t}) 
        =
        i[H,\O(\mathcal{A}_{t})]
    \end{equation}
    The equality is understood in the weak sense on \(D(H)\), i.e., for every \(\psi_i,\psi_j\in D(H)\),
    \[
        \frac{d}{dt}\langle\psi_i,\mathcal O(A_t)\psi_j\rangle_{\H}
        =
        i\left(\langle H\psi_i,\mathcal O(A_t)\psi_j\rangle_{\H}
        -
        \langle\psi_i,\mathcal O(A_t)H\psi_j\rangle_{\H}\right).
    \]
\end{Corollary}

\begin{proof}
    We first show that if \((U_t)_{t\in\mathbb R}\) is a strongly continuous one-parameter group of CFS symmetries, then the position observables satisfy the Heisenberg equation.
    
    By Stone's theorem \ref{thm:Stone}, the strongly continuous one-parameter unitary group is given by \(U_t=e^{itH}\), where \(H\) is self-adjoint. For every \(\psi\in\mathcal D(H)\), the map \(t\mapsto U_t^{-1}\psi\) is differentiable in the Hilbert space norm, with
    \[
        \frac{d}{dt}U_t^{-1}\psi
        =
        -iHU_t^{-1}\psi.
    \]
    By Theorem \ref{thm:unitary_observable}, the position observable satisfies
    \[
        \mathcal O(A_t)
        =
        U_t\mathcal O(A)U_t^{-1}.
    \]
    Since $U_t$ is a symmetry, we have $\rho(A_t)=\rho(A)<\infty$ for all $t\in\mathbb R$, so the position observables $\mathcal O(A_t)$ are well-defined. Moreover, since $\|\pi_x\|\leq 1$, the position observable is bounded, with
    \[
        \|\mathcal O(A)\|
        \leq
        \int_A\|\pi_x\|\,d\rho(x)
        \leq
        \rho(A)<\infty.
    \]
    Since $U_t$ and $U_t^{-1}$ are strongly continuous and $\mathcal O(A)$ is bounded, the covariance relation implies that $t\mapsto\mathcal O(A_t)$ is strongly continuous.
    Consequently, for arbitrary \(\psi_i,\psi_j\in\mathcal D(H)\), the boundedness of $\O(A)$ and the norm differentiability of $U_t^{-1}\psi_i$ makes the following calculation well-defined, 
    \[
    \begin{aligned}
        \frac{d}{dt}
        \langle\psi_i,\mathcal O(A_t)\psi_j\rangle_{\H}
        &=
        \frac{d}{dt}
        \langle U_t^{-1}\psi_i,\mathcal O(A)U_t^{-1}\psi_j\rangle_{\H}
        \\
        &=
        i\Big(
        \langle HU_t^{-1}\psi_i,\mathcal O(A)U_t^{-1}\psi_j\rangle_{\H}
        -
        \langle U_t^{-1}\psi_i,\mathcal O(A)HU_t^{-1}\psi_j\rangle_{\H}
        \Big)
        \\
        &=
        i\Big(
        \langle H\psi_i,\mathcal O(A_t)\psi_j\rangle_{\H}
        -
        \langle\psi_i,\mathcal O(A_t)H\psi_j\rangle_{\H}
        \Big).
    \end{aligned}
    \]
    Here we used that $\O(\mathcal{A})$ is independent of $t$, i.e.,
    $\frac{d}{dt}\O(\mathcal{A})=0$,
     and  that \(U_t\) is unitary,  preserves \(\mathcal D(H)\) and commutes with \(H\) on this domain. This establishes the forward implication.
    
     Conversely, let \(A\subset\F_n\) be a Borel set with \(\rho(A)<\infty\), and let
    \[
        A_t
        :=
        U_tAU_t^{-1},
        \qquad 
        U_t
        =
        e^{itH}.
    \]
    By assumption, the position observables \(\mathcal O(A_t)\) are well-defined as bounded operators for all \(t\in\mathbb R\), depend strongly continuously on \(t\), and satisfy the Heisenberg equation in the weak sense.
    We now show that these properties imply the covariance relation
    \[
        \mathcal O(A_t)
        =
        U_t\mathcal O(A)U_t^{-1}.
    \]
    Let \(\psi_i,\psi_j\in D(H)\). Since \(U_t\) preserves \(D(H)\) and commutes with \(H\) on this domain, the vectors \(U_t\psi_i\) are differentiable in the Hilbert space norm, with
    \[
        \frac{d}{dt}U_t\psi_i
        =
        iHU_t\psi_i.
    \]
    Moreover, the strong continuity of \(\mathcal O(A_t)\) implies local uniform boundedness in operator norm. We can therefore differentiate the matrix elements using the product rule and the assumed weak Heisenberg equation. We obtain
    \[
    \begin{aligned}
        \frac{d}{dt}
        \langle U_t\psi_i,\mathcal O(A_t)U_t\psi_j\rangle_{\H}
        &=
        -i\langle HU_t\psi_i,\mathcal O(A_t)U_t\psi_j\rangle_{\H} +i\langle U_t\psi_i,\mathcal O(A_t)HU_t\psi_j\rangle_{\H}
        \\
        &\hphantom{=}
        +i\Big(
        \langle HU_t\psi_i,\mathcal O(A_t)U_t\psi_j\rangle_{\H}
        -
        \langle U_t\psi_i,\mathcal O(A_t)HU_t\psi_j\rangle_{\H}
        \Big)
        \\
        &=
        0.
    \end{aligned}
    \]
    Consequently, the matrix elements are independent of \(t\). Evaluating at \(t=0\), where \(U_0=I\) and \(A_0=A\), gives
    \[
        \langle U_t\psi_i,\mathcal O(A_t)U_t\psi_j\rangle_{\H}
        =
        \langle\psi_i,\mathcal O(A)\psi_j\rangle_{\H}.
    \]
    By unitarity of \(U_t\), this is equivalent to
    \[
        \langle\psi_i,U_t^{-1}\mathcal O(A_t)U_t\psi_j\rangle_{\H}
        =
        \langle\psi_i,\mathcal O(A)\psi_j\rangle_{\H}.
    \]
    Since \(D(H)\) is dense in \(\H\) and both operators are bounded, we conclude that
    \[
        U_t^{-1}\mathcal O(A_t)U_t
        =
        \mathcal O(A),
    \]
    or equivalently,
    \[
        \mathcal O(A_t)
        =
        U_t\mathcal O(A)U_t^{-1}.
    \]
    Since this covariance relation holds for every Borel set \(A\subset\F_n\) with \(\rho(A)<\infty\), Theorem \ref{thm:unitary_observable} implies that \(U_t\) is a symmetry of the CFS for every \(t\in\mathbb R\). This establishes the converse implication and completes the proof.
\end{proof}

In the case of a static \ac{cfs}, Definition \ref{def:static_CFS} identifies the action of the strongly continuous one-parameter unitary group on spacetime with translations in the time coordinate. Thus, the self-adjoint generator $H$ can be interpreted as the Hamiltonian associated with this time evolution. More generally, a one-parameter unitary symmetry gives rise to a conserved quantity, namely the expectation value of its self-adjoint generator.

\begin{Lemma}
    Let $U_{t}=e^{itH}$, where $H$ is the self-adjoint generator of the strongly continuous one-parameter unitary group $\{U_{t}\}_{t\in\mathbb{R}}$. Then, for every $v\in\mathcal D(H)$, $E(v_t):=\bra v_t,Hv_t\ket$, $v_t:=U_{t}v$, is conserved under the time evolution. Equivalently,
    \[
        \frac{d}{dt}E(v_t)
        =
        0,
        \qquad
        \forall \, t \in \R.
    \]
\end{Lemma}

\begin{proof}
    Since $U_{t} = e^{itH}$ is generated by $H$, the operators $U_{t}$ and $H$ commute on $\mathcal D(H)$. Hence $U_{t}^{-1}H U_{t}=H$. Therefore,
    \begin{align*}
        E(v_t) = \bra U_{t}v,HU_{t}v\ket = \bra v,U_{t}^{-1}HU_{t}v\ket = \bra v,Hv\ket,
    \end{align*}
    which is independent of $t$. Consequently,
    \[
        \frac{d}{dt}E(v_t)
        =
        0.
    \]
\end{proof}

\subsection{Collapse Models}

Effective collapse models in \ac{cfs}  are obtained by considering stochastic perturbations of a background evolution generated by a free Hamiltonian $H_0$ \cite[Section~3.4]{Collapse}. A central assumption throughout the collapse analysis is that observables describing local measurements commute with the free Hamiltonian,
\begin{equation}
    [\O,H_0]
    =
    0,
\end{equation}
as well as with the relevant perturbation operators, reflecting the locality of the dynamics on the scale of the measurement. This motivates us to investigate under which conditions the position observables of a \ac{cfs} commute with the generator of a one-parameter group of unitary symmetries.

\begin{Lemma} \label{Commutation}
    Let $(\H,\F_n,\rho)$ be a \ac{cfs} with a strongly continuous one-parameter group of unitary symmetries $U_t=e^{itH}$, where $H$ is self-adjoint. Let $A\subset\F_n$ be a Borel set with $\rho(A)<\infty$.
    Then the position observable $\mathcal O(A)$ commutes with $H$ if either of the following conditions holds:
    \begin{enumerate}
        \item The region $A$ is invariant under the symmetry, i.e.,
    \[
        A_t
        :=
        U_tAU_t^{-1}=A
        \qquad\text{for all }
        t \in \mathbb R.
    \]
    \item The spin-space projections are invariant under the symmetry, i.e.,
    \[
        U_t\pi_xU_t^{-1}
        =
        \pi_x
    \]
    for every $t\in\mathbb R$ and $\rho$-almost every $x\in A$.
    \end{enumerate}
    Here, commutation with $H$ is understood on its domain $\mathcal D(H)$.
\end{Lemma}

\begin{proof}
    By Theorem \ref{thm:unitary_observable}, the position observable transforms as
    \[
        U_t\mathcal O(A)U_t^{-1}
        =
        \mathcal O(A_t).
    \]
    If condition a) holds, then \(A_t=A\) for all \(t\in\mathbb R\). Consequently,
    \[
        U_t\mathcal O(A)U_t^{-1}
        =
        \mathcal O(A).
    \]
    If condition b) holds, we instead obtain
    \[
    \begin{aligned}
        U_t\mathcal O(A)U_t^{-1}
        &=
        \int_A U_t\pi_xU_t^{-1}\,d\rho(x)
        =
        \int_A\pi_x\,d\rho(x)=\mathcal O(A).
    \end{aligned}
    \]
    Thus, in either case, \(\mathcal O(A)\) commutes with \(U_t\) for all \(t\in\mathbb R\). Since \(\mathcal O(A)\) is bounded and commutes with \(U_t\) for all \(t\in\mathbb R\), Stone's theorem \ref{thm:Stone} implies that \(\mathcal O(A)\) preserves \(\mathcal D(H)\) and commutes with \(H\) on this domain.
\end{proof}

Since each spin space \(S_x\) is finite-dimensional, invariance under \(U_t\) implies that \(S_x\) is spanned by eigenvectors of \(H\). Conversely, if \(S_x\) is spanned by eigenvectors of \(H\), then \(U_t\) preserves \(S_x\), and therefore commutes with \(\pi_x\).
Thus, condition b) is equivalent to requiring that \(S_x\) be spanned by eigenvectors of \(H\) for \(\rho\)-almost every \(x\in A\).

\section{Symmetry Mapping} \label{sec:symmetry_mapping}

The aim of this section is to establish a correspondence between symmetries of a spacetime and symmetries of the \ac{cfs} obtained from it through the local correlation map. In particular, we show under which assumptions spacetime isometries induce unitary symmetries of the associated \ac{cfs}. This allows us to relate the abstract generator of a continuous \ac{cfs} symmetry introduced in Section~\ref{sec:observable_&_evolution} to the explicit form of the Hamiltonian governing the Dirac evolution on static spacetimes.

To establish this correspondence, we first recall the necessary framework for the Dirac equation on globally hyperbolic spin spacetimes and introduce spin isometries together with their action on solutions to that equation. We then introduce the local correlation map and the associated construction of the \ac{cfs} measure from the spacetime measure. This provides the bridge necessary to study the mapping of symmetries between the spacetime and a particular \ac{cfs} . 

\subsection{Dirac Equation and the Cauchy Problem} \label{subsec:dirac_cauchy_problem}

We work throughout this section with a connected, globally hyperbolic Lorentzian spin manifold $(\M,g)$ of even dimension $d=2k$, $k\geq 1$, and signature $(+,-,\ldots,-)$. We denote its complex Dirac spinor bundle by
$\mathbb{S}\M$ and the corresponding bundle projection by
\[
    \pi
    :
    \mathbb{S}\M\longrightarrow \M.
\]
For every $x\in\M$, we denote the fibre over $x$ by
\[
    \mathbb{S}_x\M
    :=
    \pi^{-1}(x).
\]
Clifford multiplication is given by the bundle map
\[
    \gamma_{\M}
    :
    T\M\longrightarrow\operatorname{End}(\mathbb{S}\M),
\]
satisfying, for every $x\in\M$ and $u,v\in T_x\M$,
\begin{equation}\label{eq:Clifford_multiplication}
    \gamma_{\M} (u) \gamma_{\M} (v)
    +
    \gamma_{\M} (v) \gamma_{\M} (u)
    =
    2g(u,v)\, \mathbb{I}.
\end{equation}
Each fibre $\mathbb S_x\M$ has complex dimension $2^k$ and is equipped with a smooth, non-degenerate Hermitian form of signature $(n,n)$, where $n=2^{k-1}$,
\[
    \prec\,\cdot\mid\cdot\,\succ_x
    :
    \quad
    \mathbb S_x\M\times\mathbb S_x\M
    \longrightarrow
    \mathbb C,
\]
called the spin scalar product, with respect to which Clifford multiplication is symmetric, i.e.,
\[
    \prec
    \psi_i(x)\mid\gamma_\M(u)\psi_j(x)
    \succ_x
    =
    \prec
    \gamma_\M(u)\psi_i(x)\mid\psi_j(x)
    \succ_x
\]
for all $u\in T_x\M$ and $\psi_i,\psi_j\in\Gamma(\mathbb S\M)$. 
Here, $\Gamma(\mathbb S\M)$ denotes the space of smooth sections of the spinor bundle,
\[
    \Gamma(\mathbb{S}\M)
    := 
    \big\{\psi:\M\rightarrow\mathbb{S}\M
    \ \text{smooth}\mid \pi(\psi(x))=x
    \text{ for all }x\in\M\big\}.
\]
On this space, we consider the Dirac operator associated with the spin connection.

\begin{Def}[Dirac Equation] \label{def:Dirac_op}
    The Dirac operator $D_{\M}$ is defined as the composition of the spin connection $\nabla^{\mathbb{S}\M}$ on $\mathbb{S}\M$, induced by the Levi-Civita connection on $T\M$, and Clifford multiplication,
    \[
        D_{\M}:=i\gamma_{\M}\circ\nabla^{\mathbb{S}\M}:
        \Gamma(\mathbb{S}\M)\longrightarrow\Gamma(\mathbb{S}\M).
    \]
    Given a local orthonormal frame $\{e_\mu\}$ of $\M$, the Dirac operator takes the form
    \begin{equation}\label{eq:Dirac_operator}
        D_{\M}
        =
        i\sum_{\mu}
        \varepsilon_{\mu}\,\gamma_{\M}(e_{\mu})\nabla^{\mathbb S\M}_{e_{\mu}},
    \end{equation}
    For a mass parameter $m\geq 0$, we define the massive Dirac operator by
    \begin{equation}\label{eq:massiveDirac_operator}
        D_{\M}^{m}
        :=
        D_{\M}-m\,\mathbb{I}.
    \end{equation}
    The corresponding massive Dirac equation is
        \begin{equation}\label{eq:massiveDirac_equation}
            D_{\M}^m \psi 
            =
            0.
        \end{equation}
    In the massless case $m=0$, one has $D_{\M}^{0}=D_{\M}$, and hence the massive Dirac equation reduces to the geometric Dirac equation
    \begin{equation}\label{eq:Dirac_equation}
        D_{\M} \psi 
        =
        0.
    \end{equation}
\end{Def}

Let $\Sigma\subset\M$ be a spacelike Cauchy surface. We denote the restriction of the spinor bundle to $\Sigma$ by
\[
    \left.\mathbb S\M\right|_\Sigma
    :=
    \pi^{-1}(\Sigma)
    =
    \bigcup_{x\in\Sigma}\mathbb S_x\M.
\]
We further denote by
\[
    \Gamma_c\!\left(\left.\mathbb S\M\right|_\Sigma\right)
\]
the space of smooth compactly supported sections of the restricted spinor bundle.

\begin{Def}[Restriction Map]
    Let $\Sigma\subset\M$ be a spacelike Cauchy surface. The restriction map is defined by
    \[
        R_\Sigma:
        \Gamma(\mathbb S\M)
        \longrightarrow
        \Gamma\!\left(\left.\mathbb S\M\right|_\Sigma\right),
        \qquad
        R_\Sigma\psi:=\left.\psi\right|_\Sigma .
    \]
\end{Def}

Using the restriction map, we denote the space of smooth spatially compact sections of the spinor bundle by
\[
    \Gamma_{\mathrm{sc}}(\mathbb S\M)
    :=
    \left\{
    \psi\in\Gamma(\mathbb S\M)
    \;\middle|\;
    R_\Sigma\psi
    \in
    \Gamma_c\!\left(\left.\mathbb S\M\right|_\Sigma\right)
    \text{ for every spacelike Cauchy surface }\Sigma
    \right\}.
\]
Thus, a smooth spinor field is spatially compact if its restriction to every spacelike Cauchy surface has compact support.
We can now formulate the Cauchy problem for the Dirac equation. Fix a mass parameter $m\geq 0$, and denote the space of smooth spatially compact solutions by
\[
    \Gamma_{\mathrm{sc}}^{m}(\mathbb S\M)
    :=
    \left\{
    \psi\in\Gamma_{\mathrm{sc}}(\mathbb S\M)
    \;\middle|\;
    D_{\M}^{m}\psi=0
    \right\}.
\]
Let $\Sigma\subset\M$ be a spacelike Cauchy surface and let
\[
    \psi_\Sigma\in\Gamma_c(\mathbb S\M|_\Sigma)
\]
be smooth compactly supported initial data. The Cauchy problem for the massive Dirac equation consists of finding
\[
    \psi\in\Gamma_{\mathrm{sc}}^{m}(\mathbb S\M)
\]
such that
\[
    R_\Sigma\psi=\psi_\Sigma.
\]
On a globally hyperbolic spacetime, the Dirac Cauchy problem is well-posed: for every
\[
    \psi_\Sigma\in\Gamma_c(\mathbb S\M|_\Sigma)
\]
there exists a unique
\[
    \psi\in\Gamma_{\mathrm{sc}}^{m}(\mathbb S\M)
\]
with $R_\Sigma\psi=\psi_\Sigma$ \cite[Theorem 2.8]{Dirac_Operator-2}. Hence, the restriction map
\[
    R_\Sigma^m
    :=
    R_\Sigma\big|_{\Gamma_{\mathrm{sc}}^{m}(\mathbb S\M)}: 
    \Gamma_{\mathrm{sc}}^{m}(\mathbb S\M)
    \longrightarrow
    \Gamma_c(\mathbb S\M|_\Sigma).
\]
is a linear bijection. Its inverse is the solution map 
\[
    (R_\Sigma^m)^{-1}:
    \Gamma_c(\mathbb S\M|_\Sigma)
    \longrightarrow
    \Gamma_{\mathrm{sc}}^{m}(\mathbb S\M)
\]
identifying every Cauchy datum $\psi_\Sigma$ with a unique solution $\psi$ satisfying
$D_\M^m\psi=0$. 

For two smooth spinor fields
$\psi_i,\psi_j\in\Gamma(\mathbb{S}\M)$, the components of the Dirac current are defined as in \cite[Equation 1.34]{CFS_Book} by
\[
    J^{\mu}_{i,j}(x)
    :=
    \varepsilon_\mu\prec\psi_i(x)\mid\gamma_\M(e_\mu)\psi_j(x)\succ_x, 
\]
where $\{e_\mu\}$ is a local pseudo-orthonormal frame field of $T\M$ and $\varepsilon_\mu\in\{+1,-1\}$ defined by 
$g(e_\mu,e_\nu)=\varepsilon_\mu\delta_{\mu\nu}$. The current vector is then given by $J_{i,j}=J^\mu_{i,j}e_\mu$. Since the spin scalar product is sesquilinear and Clifford multiplication is linear, the Dirac current is anti-linear in $\psi_i$ and linear in $\psi_j$. Moreover, symmetry of Clifford multiplication with respect to the spin scalar product gives
\[
    J^{\mu}_{i,j}(x)
    =
    \overline{J^{\mu}_{j,i}(x)}.
\]
Let now $\Sigma\subset\M$ be a spacelike Cauchy surface, with future-directed unit normal vector field $\nu_\Sigma$. Since $\nu_\Sigma(x)$ is timelike, the operator $\gamma_\M(\nu_\Sigma(x))$ turns the spin scalar product into a positive-definite Hermitian form on $\mathbb S_x\M$. Thus, following \cite[Equation 1.37]{CFS_Book} for
\[
    \psi_i,\psi_j\in\Gamma_c(\mathbb S\M|_\Sigma),
\]
the normal component of the Dirac current induces the scalar product
\begin{equation}\label{eq:scalar_product}
    \bra\psi_i,\psi_j\ket_{\Sigma}
    :=
    \int_\Sigma
    \prec\!\psi_i(x)
    \!\mid\!
    \gamma_{\M}(\nu_\Sigma)\psi_j(x)
    \!\succ_x
    d\mu_\Sigma(x),
\end{equation}
where $d\mu_\Sigma$ is the metric volume measure induced on $\Sigma$.

\begin{Def}[Hypersurface Hilbert Space]\label{def:hypersurface_H_space}
    Let $\bra \cdot,\cdot \ket_{\Sigma}$ be the scalar product on $\Gamma_{c}(\mathbb{S}\M|_{\Sigma})$ defined by Equation \eqref{eq:scalar_product}. The hypersurface Hilbert space is defined as the completion
    \[
        \H_{\Sigma}
        :=
        \overline{\Gamma_{c}(\mathbb{S}\M|_{\Sigma})}^{\bra \cdot,\cdot \ket_{\Sigma}} \, .
    \]
\end{Def}

For solutions of the Dirac equation, the Dirac current satisfies the conservation law given in \cite[Equations 1.35 and 1.36]{CFS_Book}, yielding the following proposition.

\begin{Prp} \label{prp:current_conservation} 
    Let $m\geq0$ and
    \[
        \psi_i,\psi_j\in\Gamma_{\mathrm{sc}}^{m}(\mathbb S\M).
    \]
    Then the Dirac current $J_{i,j}=J^\mu_{i,j}e_\mu$ is divergence-free,
    \[
        \nabla_\mu J^\mu_{i,j}=0.
    \]
    Consequently, for any two spacelike Cauchy surfaces $\Sigma,\Sigma'\subset\M$, with corresponding future-directed timelike unit normal vector fields $\nu_\Sigma$, $\nu_{\Sigma'}$
    \[
        \int_\Sigma
        \prec\psi_i(x)\mid\gamma_\M(\nu_\Sigma(x))\psi_j(x)\succ_x\,d\mu_\Sigma(x)
        =
        \int_{\Sigma'}
        \prec\psi_i(x)\mid\gamma_\M(\nu_{\Sigma'}(x))\psi_j(x)\succ_x\,d\mu_{\Sigma'}(x).
    \]
\end{Prp}

Let $\Sigma\subset \M$ be a spacelike Cauchy surface. The restriction map $R_\Sigma^m$ allows us to define a scalar product on the space on $\Gamma_{\mathrm{sc}}^{m}(\mathbb S \M)$ by
\begin{equation} \label{eq:solution_scalar_product}
    \bra\psi_i,\psi_j\ket_{m}
    :=
    \left\bra
    R_\Sigma^m\psi_i,
    R_\Sigma^m\psi_j
    \right\ket_{\Sigma}
\end{equation}
By Proposition \ref{prp:current_conservation}, this scalar product is independent of the choice of the spacelike Cauchy surface $\Sigma$.

\begin{Def}[Dirac Hilbert Space] \label{def:Sol_Hilbert_space}
    Let $\bra\cdot,\cdot\ket_{m}$ be the scalar product on $\Gamma_{\mathrm{sc}}^{m}(\mathbb S \M)$ defined by Equation \eqref{eq:solution_scalar_product}. The Dirac Hilbert space is defined as the completion
    \[
        \H^m
        :=
        \overline{
        \Gamma_{\mathrm{sc}}^{m}(\mathbb S \M)
        }^{\bra\cdot,\cdot\ket_{m}}.
    \]
\end{Def}

We have thus constructed the Dirac Hilbert space from the space of smooth spatially compact solutions. The well-posedness of the Cauchy problem gives a bijection between these smooth solutions and smooth compactly supported Cauchy data. The following proposition shows that this correspondence extends to the respective Hilbert-spaces.

\begin{Prp} \label{prop:restriction_unitary}
    Let $m\geq 0$ and let $\Sigma\subset \M$ be a spacelike Cauchy surface. The restriction map
    \[
        R_\Sigma^m:
        \Gamma_{\mathrm{sc}}^{m}(\mathbb S \M)
        \longrightarrow
        \Gamma_c(\mathbb S \M|_\Sigma)
    \]
    extends uniquely to a unitary map
    \[
        R_\Sigma^m:
        \H^m
        \longrightarrow
        \H_\Sigma.
    \]
    Its inverse is the unique continuous extension of the solution map,
    \[
        (R_\Sigma^m)^{-1}:
        \H_\Sigma
        \longrightarrow
        \H^m.
    \]
\end{Prp}

\begin{proof}
    By the definition of the solution scalar product in Equation \eqref{eq:solution_scalar_product}, $R_\Sigma^m$ is an isometry on $ \Gamma_{\mathrm{sc}}^{m} (\mathbb S \M) $. It therefore extends uniquely by continuity to an isometry
    \[
        R_\Sigma^m:
        \H^m
        \longrightarrow
        \H_\Sigma.
    \]
    The range of an isometry between Hilbert spaces is closed. Moreover, by well-posedness of the Cauchy problem, the smooth restriction map is surjective onto $\Gamma_c(\mathbb S \M|_\Sigma)$, which is dense in $\H_\Sigma$. Hence the range of the extended map is both closed and dense in $\H_\Sigma$, and therefore equals $\H_\Sigma$. Thus $R_\Sigma^m$ is unitary. Its inverse is consequently unitary and agrees on $\Gamma_c(\mathbb S \M|_\Sigma)$ with the solution map $(R_\Sigma^m)^{-1}$. It is therefore the unique continuous extension of the solution map to $\H_\Sigma$.
\end{proof}

\begin{Remark}
    In the following, whenever a map between the smooth spaces admits a unique continuous extension to the corresponding Hilbert-spaces, we denote the continuous extension by the same symbol as the original map.
\end{Remark}

The restriction maps allow us to identify the Cauchy data on different spacelike Cauchy surfaces via the following evolution map.

\begin{Corollary}\label{cor:evolution_unitary}
    Let $\Sigma$ and $\Sigma'$ be Cauchy surfaces. Then the evolution map
    \[
        E^m_{\Sigma',\Sigma} : \H_{\Sigma} \longrightarrow\H_{\Sigma'},
        \qquad
        E^m_{\Sigma',\Sigma}
        :=
        R^m_{\Sigma'} \left(R^m_{\Sigma}\right)^{-1}
    \]
    between the two Cauchy surfaces is unitary and
    \[        
        E^m_{\Sigma',\Sigma}
        =
        \left(E^m_{\Sigma,\Sigma'}\right)^{-1}.
    \]
\end{Corollary}

\begin{proof}
    Since $R^m_{\Sigma}$ is unitary for all $\Sigma$, the composition $E^m_{\Sigma',\Sigma}
        =
        R^m_{\Sigma'} \left(R^m_{\Sigma}\right)^{-1}$ is unitary. Moreover,
    \[        
        E^m_{\Sigma',\Sigma}
        =
        R^m_{\Sigma'} \left(R^m_{\Sigma}\right)^{-1}
        =
        (R^m_{\Sigma} \left(R^m_{\Sigma'}\right)^{-1})^{-1}
        =
        \left(E^m_{\Sigma,\Sigma'}\right)^{-1}.
    \]
\end{proof}

The relation between the Dirac Hilbert space and the hypersurface Hilbert spaces associated with different Cauchy surfaces, together with their unitary identifications, is summarized by the following commutative diagram:
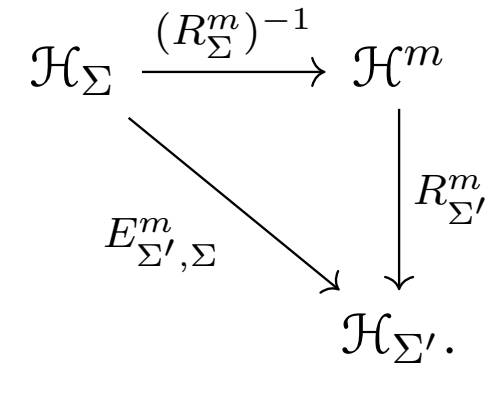
\[
    \begin{tikzcd}[column sep=3em, row sep=3em]
        \H_\Sigma
            \arrow[r, "(R^m_{\Sigma})^{-1}"]
            \arrow[dr, swap, "E^m_{\Sigma',\Sigma}"]
        &
        \H^m
            \arrow[d, "R^m_{\Sigma'}"]
        \\
        &
        \H_{\Sigma'} .
    \end{tikzcd}
\]

\subsection{Isometry-Induced Unitary Maps} \label{subsec:general_spin_isometries}

Having established the relation between the Dirac Hilbert space and the hypersurface Hilbert spaces, we now study how symmetries of the spin geometry act on these spaces. We begin with the ordinary notion of an isometry of spacetime.

\begin{Def}[Isometry] \label{def:spacetime_isometry_action}
    Let $(\M,g)$ be a Lorentzian manifold. A diffeomorphism
    \[
        \Phi:\M\longrightarrow \M
    \]
    is called an isometry if
    \[
        \Phi^*g=g.
    \]
    Thus, for every $x\in \M$ and $u,v\in T_x\M$,
    \[
        g_x(u,v)= g_{\Phi(x)}\bigl(d\Phi_x(u),d\Phi_x(v)\bigr)
    \]
    where $ d\Phi_x : T_x \M \longrightarrow T_{\Phi(x)} \M$ denotes the differential map of $\Phi$ at $x$.
\end{Def}

In the spin-geometric setting, this notion is complemented by that of a spin isometry.

\begin{Def}[Spin Isometry] \label{def:spin_isometry}
    Let $(\M,g)$ be a Lorentzian spin manifold with spinor bundle $\mathbb S \M$ and bundle projection $\pi:\mathbb S \M\to \M$. We consider a bundle map $\widetilde{\Phi}:\mathbb S \M\to\mathbb S \M$ covering a diffeomorphism $\Phi:\M\to \M$, i.e.,
    \[
        \pi\circ\widetilde{\Phi}=\Phi\circ\pi
        \qquad\text{as maps }\mathbb S \M\longrightarrow \M.
    \]
    Its restriction to each fibre defines a linear map
    \[
        \left.\widetilde{\Phi}\right|_x:
        \mathbb S_x\M\longrightarrow\mathbb S_{\Phi(x)}\M,
        \qquad x\in \M.
    \]
    The bundle map $\widetilde{\Phi}$ is called a spin isometry if it preserves the spin scalar product and is compatible with Clifford multiplication. More precisely, for all $x\in \M$, $\psi_i,\psi_j\in\Gamma(\mathbb S \M)$, and $u\in T_x\M$, the following relations hold:
    \begin{equation} \label{eq:isometryspinproduct}
        \prec
        \left.\widetilde{\Phi}\right|_x\bigl(\psi_i(x)\bigr)
        \mid
        \left.\widetilde{\Phi}\right|_x\bigl(\psi_j(x)\bigr)
        \succ_{\Phi(x)}
        =
        \prec\psi_i(x)\mid\psi_j(x)\succ_x,
    \end{equation}
    and
    \begin{equation}\label{eq:isometry_and_Clifford_multiplication}
        \left.\widetilde{\Phi}\right|_x
        \bigl(\gamma_\M(u)\psi(x)\bigr)
        =
        \gamma_\M\bigl(d\Phi_x(u)\bigr)
        \left.\widetilde{\Phi}\right|_x\bigl(\psi(x)\bigr).
    \end{equation}
\end{Def}

Our definition differs slightly from the standard convention, where the covered diffeomorphism is assumed to be an isometry from the outset, e.g. in  \cite[Definition~3.1]{SigSymm}. This assumption is not needed, since compatibility with Clifford multiplication already implies that the covered diffeomorphism is an isometry, as shown in the following proposition. Thus, the two definitions agree.

\begin{Prp} \label{prp:spin_implies_isometry}
    Let $(\M,g)$ be a Lorentzian spin manifold and let $\widetilde{\Phi}:\mathbb S \M\to\mathbb S \M$ be a spin isometry covering a diffeomorphism $\Phi:\M\to \M$. Then $\Phi$ is an isometry of $(\M,g)$.
\end{Prp}

\begin{proof}
    Preservation of the non-degenerate spin scalar product implies that each fibre map
    \[
        \left.\widetilde{\Phi}\right|_x
        :
        \mathbb S_x\M\longrightarrow\mathbb S_{\Phi(x)}\M
    \]
    is injective and hence an isomorphism. Compatibility with Clifford multiplication therefore gives
    \[
        \gamma_\M\bigl(d\Phi_x(u)\bigr)
        =
        \left.\widetilde{\Phi}\right|_x
        \gamma_\M(u)
        \left(\left.\widetilde{\Phi}\right|_x\right)^{-1} \text{ in } 
        \operatorname{End}\bigl(\mathbb S_{\Phi(x)}\M\bigr)\,.
    \]
    Using the Clifford relation, for $u,v\in T_x\M$ we obtain
    \[
        2g_{\Phi(x)}\bigl(d\Phi_x(u),d\Phi_x(v)\bigr)\mathbb I_{\mathbb S_{\Phi(x)}\M}
        =
        \left.\widetilde{\Phi}\right|_x
        \bigl(2g_x(u,v)\mathbb I_{\mathbb S_x\M}\bigr)
        \left(\left.\widetilde{\Phi}\right|_x\right)^{-1}
        =
        2g_x(u,v)\mathbb I_{\mathbb S_{\Phi(x)}\M}.
    \]
    Hence
    \[
        g_{\Phi(x)}\bigl(d\Phi_x(u),d\Phi_x(v)\bigr)=g_x(u,v),
    \]
    and $\Phi$ is an isometry.
\end{proof}

Next we return to the globally hyperbolic setting to describe how a spin isometry acts on Cauchy data. This gives a geometric identification of the hypersurface Hilbert spaces associated with $\Sigma$ and $\Phi(\Sigma)$, which will subsequently be used to establish unitarity of the induced action on the Dirac Hilbert space.

\begin{Def}[Geometric Flow Map] \label{def:geometric_flow_map}
    Let $(\M,g)$ be a globally hyperbolic Lorentzian spin manifold, let $\widetilde{\Phi}:\mathbb S \M\to\mathbb S \M$ be a spin isometry covering $\Phi:\M\to \M$, and let $\Sigma\subset \M$ be a spacelike Cauchy surface. The geometric flow map is the linear map
    \[
        \G_{\Phi(\Sigma),\Sigma}
        :
        \Gamma_{c}(\mathbb{S}\M|_\Sigma)
        \longrightarrow
        \Gamma_{c}\big(
        \mathbb{S}\M|_{\Phi(\Sigma)}
        \big)
    \]
    defined by,
    \begin{equation} \label{eq:cauchy_spin_isometry_action}
        (\G_{\Phi(\Sigma), \Sigma}\, \psi_{\Sigma}) (\Phi(x))
        :=
        \widetilde{\Phi} \big|_x
        \big(
        \psi_{\Sigma}(x)
        \big),
        \qquad 
        \psi_{\Sigma} \in \Gamma_{c}(\mathbb{S}\M|_{\Sigma}).
    \end{equation}
\end{Def}

\begin{Prp} \label{prop:cauchy_spin_isometry_unitary}
    Let $(\M,g)$ be a globally hyperbolic Lorentzian spin manifold, let $\widetilde{\Phi} : \mathbb S \M\longrightarrow\mathbb S \M$ be a spin isometry covering a time-orientation-preserving diffeomorphism $\Phi:\M\to \M$, and let $\Sigma\subset \M$ be a spacelike Cauchy surface. Then the geometric flow map extends uniquely to a unitary map
    \[
        \G_{\Phi(\Sigma),\Sigma}:
        \H_\Sigma\longrightarrow\H_{\Phi(\Sigma)}.
    \]
\end{Prp}

\begin{proof}
    Since $\Phi$ is a time-orientation-preserving isometry, it maps the future-directed unit normal of $\Sigma$ to the future-directed unit normal of $\Phi(\Sigma)$, i.e.,
    \begin{equation}\label{eq:normalvector}
        \nu_{\Phi(\Sigma)}(\Phi(x))
        =
        d\Phi_x\bigl(\nu_\Sigma(x)\bigr).  
    \end{equation}
    Moreover, since the hypersurface measures are induced by the metric and $\Phi$ is an isometry,
    \begin{equation}\label{eq:HSvolumemeasure}
        (\Phi|_\Sigma)^*d\mu_{\Phi(\Sigma)}
        =
        d\mu_\Sigma. 
    \end{equation}
    Now let
    $\psi_i,\psi_j\in\Gamma_c(\mathbb S \M|_\Sigma)$.
    By definition of the hypersurface scalar product,
    \[
    \begin{aligned}
        &\left\bra
        \G_{\Phi(\Sigma),\Sigma}\psi_i,
        \G_{\Phi(\Sigma),\Sigma}\psi_j
        \right\ket_{\Phi(\Sigma)}
        \\
        &\qquad=
        \int_{\Phi(\Sigma)}
        \prec
        \bigl(\G_{\Phi(\Sigma),\Sigma}\psi_i\bigr)(y)
        \mid
        \gamma_\M\bigl(\nu_{\Phi(\Sigma)}(y)\bigr)
        \bigl(\G_{\Phi(\Sigma),\Sigma}\psi_j\bigr)(y)
        \succ_y\,
        d\mu_{\Phi(\Sigma)}(y).
    \end{aligned}
    \]
    Making the change of variables $y=\Phi(x)$ and using \eqref{eq:normalvector} and \eqref{eq:HSvolumemeasure}, we obtain
    \[
    \begin{aligned}
        &\left\bra
        \G_{\Phi(\Sigma),\Sigma}\psi_i,
        \G_{\Phi(\Sigma),\Sigma}\psi_j
        \right\ket_{\Phi(\Sigma)}
        \\
        &\qquad=
        \int_\Sigma
        \prec
        \left.\widetilde{\Phi}\right|_x\bigl(\psi_i(x)\bigr)
        \mid
        \gamma_\M\bigl(d\Phi_x(\nu_\Sigma(x))\bigr)
        \left.\widetilde{\Phi}\right|_x\bigl(\psi_j(x)\bigr)
        \succ_{\Phi(x)}
        \,d\mu_\Sigma(x).
    \end{aligned}
    \]
    By compatibility with Clifford multiplication,
    \[
        \gamma_\M\bigl(d\Phi_x(\nu_\Sigma(x))\bigr)
        \left.\widetilde{\Phi}\right|_x\bigl(\psi_j(x)\bigr)
        =
        \left.\widetilde{\Phi}\right|_x
        \bigl(\gamma_\M(\nu_\Sigma(x))\psi_j(x)\bigr).
    \]
    Preservation of the spin scalar product therefore gives
    \[
    \begin{aligned}
        \left\bra
        \G_{\Phi(\Sigma),\Sigma}\psi_i,
        \G_{\Phi(\Sigma),\Sigma}\psi_j
        \right\ket_{\Phi(\Sigma)}
        &=
        \int_\Sigma
        \prec
        \psi_i(x)
        \mid
        \gamma_\M(\nu_\Sigma(x))\psi_j(x)
        \succ_x\,d\mu_\Sigma(x)
        \\
        &=
        \bra\psi_i,\psi_j\ket_\Sigma.
    \end{aligned}
    \]
    Thus $\G_{\Phi(\Sigma),\Sigma}$ is an isometry between $\Gamma_{c}(\mathbb{S}\M|_\Sigma)$ and $\Gamma_{c}(\mathbb{S}\M|_{\Phi(\Sigma)})$. Moreover, its inverse is induced by the inverse spin isometry $\widetilde{\Phi}^{-1}$, so it is bijective on these spaces. The same completion argument as in the proof of Proposition \ref{prop:restriction_unitary} therefore shows that $\G_{\Phi(\Sigma),\Sigma}$ extends uniquely to a unitary map
    \[
        \G_{\Phi(\Sigma),\Sigma}:
        \H_\Sigma\longrightarrow\H_{\Phi(\Sigma)}.
    \]
\end{proof}

A spin isometry $\widetilde{\Phi}$ also acts naturally on smooth sections of the spinor bundle. For $\psi\in\Gamma(\mathbb S \M)$, the section $\widetilde{\Phi}\psi\in\Gamma(\mathbb S \M)$ is defined by
\begin{equation}\label{eq:spinaction}
    (\widetilde{\Phi}\psi)(\Phi(x))
    :=
    \left.\widetilde{\Phi}\right|_x\bigl(\psi(x)\bigr),
    \qquad x\in \M.
\end{equation}
This action preserves spatial compactness. Indeed, for every spacelike Cauchy surface $\Sigma$,
\[
    R_{\Phi(\Sigma)}(\widetilde{\Phi}\psi)
    =
    \G_{\Phi(\Sigma),\Sigma}R_\Sigma\psi.
\]
Hence, if $\psi\in\Gamma_{\mathrm{sc}}(\mathbb S \M)$, then
\[
    R_{\Phi(\Sigma)}(\widetilde{\Phi}\psi)
    \in
    \Gamma_c(\mathbb S \M|_{\Phi(\Sigma)}).
\]
Since $\Phi$ is an isometry, it maps spacelike Cauchy surfaces bijectively to spacelike Cauchy surfaces. It follows that
\[
    \widetilde{\Phi}:
    \Gamma_{\mathrm{sc}}(\mathbb S \M)
    \longrightarrow
    \Gamma_{\mathrm{sc}}(\mathbb S \M)
\]
is a bijection. To study the action of spin isometries on solutions of the Dirac equation, we require an additional assumption.

\begin{Def}[Dirac-Compatible Spin Isometries] \label{def:solution_spin_map}
    Let $(\M,g)$ be a globally hyperbolic Lorentzian spin manifold. We call a spin isometry $\widetilde{\Phi}:\mathbb S \M\to\mathbb S \M$ Dirac compatible if it maps the kernel of the massive Dirac operator bijectively onto itself, i.e.,
    \[ \widetilde{\Phi}\bigl(\ker(D_\M^m|_{\Gamma_{\mathrm{sc}}(\mathbb S \M)})\bigr) = \ker(D_\M^m|_{\Gamma_{\mathrm{sc}}(\mathbb S \M)}). \]
    We define the induced map on the solution space as
    \[ 
        U^m
        :=
        \left.\widetilde{\Phi}\right|_{\Gamma_{\mathrm{sc}}^m(\mathbb S \M)} 
        : 
        \Gamma_{\mathrm{sc}}^m(\mathbb S \M) \longrightarrow \Gamma_{\mathrm{sc}}^m(\mathbb S \M). 
    \]
\end{Def}

Combining Dirac compatibility with the time-orientation-preserving assumption of Proposition \ref{prop:cauchy_spin_isometry_unitary}, we obtain the following result (see also \cite[Lemma~3.4 and Eq.~(3.7)]{SigSymm}).

\begin{Prp}\label{prop:solution_spin_isometry_unitary}
    Let $(\M,g)$ be a globally hyperbolic Lorentzian spin manifold, and let $\widetilde{\Phi}:\mathbb S \M\to\mathbb S \M$ be a Dirac-compatible spin isometry covering a time-orientation-preserving diffeomorphism $\Phi:\M\to \M$. Then the induced map
    \[
        U^m:\Gamma_{\mathrm{sc}}^m(\mathbb S \M)
        \longrightarrow
        \Gamma_{\mathrm{sc}}^m(\mathbb S \M)
    \]
    extends uniquely to a unitary map
    \[
        U^m:\H^m\longrightarrow\H^m.
    \]
\end{Prp}

\begin{proof}
    Let $\psi_i,\psi_j\in\Gamma_{\mathrm{sc}}^m(\mathbb S \M)$, and let $\Sigma\subset \M$ be a spacelike Cauchy surface. By Definitions \ref{def:geometric_flow_map} and \ref{def:solution_spin_map}, for every $x\in\Sigma$ and $\psi\in\Gamma_{\mathrm{sc}}^m(\mathbb S \M)$, we have
    \[
    \begin{aligned}
        \bigl(R_{\Phi(\Sigma)}^m U^m\psi\bigr)(\Phi(x))=(U^m\psi)(\Phi(x))=\left.\widetilde\Phi\right|_x\bigl(\psi(x)\bigr)=\bigl(\G_{\Phi(\Sigma),\Sigma}R_\Sigma^m\psi\bigr)(\Phi(x)).
    \end{aligned}
    \]
    Consequently,
    \[
        R_{\Phi(\Sigma)}^m U^m
        =
        \G_{\Phi(\Sigma),\Sigma}R_\Sigma^m.
    \]
    Using the definition of the scalar product \eqref{eq:solution_scalar_product} on $\Phi(\Sigma)$ for the transformed solutions, we obtain
    \[
    \begin{aligned}
        \bra U^m\psi_i,U^m\psi_j\ket_m
        &=\left\bra
        R_{\Phi(\Sigma)}^m U^m\psi_i,
        R_{\Phi(\Sigma)}^m U^m\psi_j
        \right\ket_{\Phi(\Sigma)}
        \\
        &=\left\bra
        \G_{\Phi(\Sigma),\Sigma}R_\Sigma^m\psi_i,
        \G_{\Phi(\Sigma),\Sigma}R_\Sigma^m\psi_j
        \right\ket_{\Phi(\Sigma)}
        =\left\bra
        R_\Sigma^m\psi_i,
        R_\Sigma^m\psi_j
        \right\ket_\Sigma
        =\bra\psi_i,\psi_j\ket_m.
    \end{aligned}
    \]
    The third equality follows from Proposition \ref{prop:cauchy_spin_isometry_unitary}, while the last equality is the definition of the scalar product \eqref{eq:solution_scalar_product} on $\Sigma$.
    
    Thus, $U^m$ is an isometry on $\Gamma_{\mathrm{sc}}^m(\mathbb S \M)$. By Dirac compatibility, $U^m$ is bijective on this space. The same completion argument as in Proposition \ref{prop:restriction_unitary} therefore shows that $U^m$ extends uniquely to a unitary operator on $\H^m$.
\end{proof}

\begin{Def}[Hypersurface Unitary] \label{def:hypersurface_unitary}
    Under the assumptions of Proposition \ref{prop:solution_spin_isometry_unitary}, let $\Sigma\subset \M$ be a spacelike Cauchy surface. We define the hypersurface unitary associated with $\widetilde\Phi$ on $\Sigma$ by
    \[
        U_\Sigma^m
        :=
        R_\Sigma^m U^m(R_\Sigma^m)^{-1}
        :
        \H_\Sigma\longrightarrow\H_\Sigma.
    \]
\end{Def}

Using the compatibility of $U^m$ with the restriction to Cauchy surfaces, we obtain
\[
\begin{alignedat}{2}
    U_\Sigma^m
    &=R_\Sigma^m U^m(R_\Sigma^m)^{-1}
    =R_\Sigma^m(R_{\Phi(\Sigma)}^m)^{-1}\G_{\Phi(\Sigma),\Sigma}
    &&=E^m_{\Sigma,\Phi(\Sigma)}\G_{\Phi(\Sigma),\Sigma}\\
    &=R_\Sigma^m U^m(R_{\Phi^{-1}(\Sigma)}^m)^{-1}R_{\Phi^{-1}(\Sigma)}^m(R_\Sigma^m)^{-1}
    &&=\G_{\Sigma,\Phi^{-1}(\Sigma)}E^m_{\Phi^{-1}(\Sigma),\Sigma}.
\end{alignedat}
\]
Together with the relation
\begin{equation*} 
    \G_{\Phi(\Sigma),\Sigma}= R^{m}_{\Phi(\Sigma)}
    U^m(
    R^{m}_\Sigma)^{-1} .
\end{equation*}
established in Proposition \ref{prop:solution_spin_isometry_unitary}, these identities can be represented by the following three commutative diagrams.
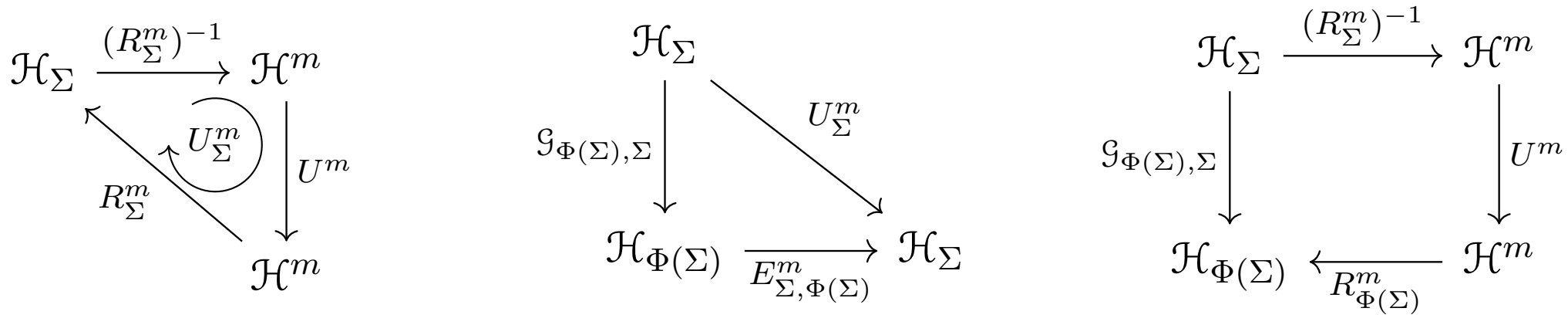
\begin{figure}[H]
\centering
\begin{minipage}{0.31\textwidth}
\centering
\begin{tikzcd}[
    column sep=3em,
    row sep=3em,
    execute at end picture={
        \coordinate (C) at
            ($ (A.center)!0.55!(B.center)!0.35!(D.center) $);
        \draw[->] ($(C)+(120:4mm)$)
            arc[start angle=120,end angle=-180,radius=4mm];
        \node at (C) {\scriptsize $U_\Sigma^m$};
    }
]
    |[alias=A]| \H_\Sigma
        \arrow[r, "(R_\Sigma^m)^{-1}"]
    &
    |[alias=B]| \H^m
        \arrow[d, "U^m"]
    \\
    &
    |[alias=D]| \H^m
        \arrow[ul, "R_\Sigma^m"]
\end{tikzcd}
\end{minipage}%
\hfill
\begin{minipage}{0.31\textwidth}
\centering
\begin{tikzcd}[column sep=3em, row sep=3em]
    \H_\Sigma
        \arrow[d, "\G_{\Phi(\Sigma),\Sigma}"']
        \arrow[dr, "U_\Sigma^m"]
    &
    \\
    \H_{\Phi(\Sigma)}
        \arrow[r, "E^m_{\Sigma,\Phi(\Sigma)}"']
    &
    \H_\Sigma
\end{tikzcd}
\end{minipage}%
\hfill
\begin{minipage}{0.31\textwidth}
\centering
\begin{tikzcd}[column sep=3em, row sep=3em]
    \H_\Sigma
        \arrow[r, "(R_\Sigma^m)^{-1}"]
        \arrow[d, "\G_{\Phi(\Sigma),\Sigma}"']
    &
    \H^m
        \arrow[d, "U^m"]
    \\
    \H_{\Phi(\Sigma)}
    &
    \H^m
        \arrow[l, "R_{\Phi(\Sigma)}^m"]
\end{tikzcd}
\end{minipage}
\end{figure}
All the maps introduced in Sections \ref{subsec:dirac_cauchy_problem} and \ref{subsec:general_spin_isometries}, together with their domains, codomains, and geometric interpretations, are summarized in Table \ref{tab:maps_summary}. For the unitary maps induced by spin isometries, we assume throughout that the underlying spacetime isometries preserve time orientation.
\begin{table}[H]
\centering
    \begin{tabular}{c c p{8cm}}
    
        \hline
        
        \textbf{Map} & \textbf{Mapping} & \textbf{Interpretation} \\
        
        \hline

        $\Phi$ & $\M \longrightarrow \M$ 
        & Spacetime isometry. \\[0.4em]

        $\widetilde{\Phi}$ & $\mathbb{S}\M \longrightarrow \mathbb{S}\M$ 
        & Spin isometry covering $\Phi$. \\[0.4em]
        
        $R^{m}_\Sigma$ & $\H^{m} \longrightarrow \H_\Sigma$ 
        & Restriction of Dirac solutions to the Cauchy surface $\Sigma$. \\[0.4em]
        
        $(R^{m}_\Sigma)^{-1}$ & $\H_\Sigma \longrightarrow \H^{m}$ 
        & Reconstruction of a global Dirac solution from Cauchy data. \\[0.4em]
        
        $E^{m}_{\Sigma',\Sigma}$ & $\H_\Sigma \longrightarrow \H_{\Sigma'}$ 
        & Cauchy evolution from $\Sigma$ to $\Sigma'$. \\[0.4em]

        $\mathcal{G}_{\Phi(\Sigma),\Sigma}$ 
        & $\H_\Sigma \longrightarrow \H_{\Phi(\Sigma)}$ 
        & Geometric transport of Cauchy data induced by $\widetilde{\Phi}$. \\[0.4em]
        
        $U^m$ & $\H^{m} \longrightarrow \H^{m}$ 
        & Unitary action of a Dirac-compatible spin isometry on solutions. \\[0.4em]

        $U^m_\Sigma$ & $\H_{\Sigma} \longrightarrow \H_{\Sigma}$ 
        & Induced unitary action on the fixed hypersurface Hilbert space. \\
        
        \hline
        
    \end{tabular}
    \caption{Summary of the maps introduced in Sections \ref{subsec:dirac_cauchy_problem} and \ref{subsec:general_spin_isometries}.}
    \label{tab:maps_summary}
\end{table}

\subsection{Mapping to Causal Fermion Systems} \label{sec:spacetime_symmetry_mapping_to_cfs}

With this understanding of spinor bundle symmetries at hand, we next establish the bridge that allows us to map spacetime structures into the framework of \ac{cfs}, namely the local correlation map.
To construct this map, we begin with the pointwise spinor scalar product, which gives rise to a family of sesquilinear forms
\begin{equation} \label{eq:Bx}
    B_x(\psi_i,\psi_j):=\prec\psi_i(x)\mid\psi_j(x)\succ_x\,,
    \qquad x\in\M,
\end{equation}
defined for smooth solutions $\psi_i,\psi_j\in(\Gamma(\mathbb S \M)\cap \H^m)$ of the Dirac equation\footnote{In the construction of \ac{cfs} from the Dirac equation, a regularization is generally necessary because pointwise evaluation of solutions is not bounded with respect to the Dirac Hilbert space norm. One possible regularization consists of introducing an ultraviolet cutoff, restricting the admissible solutions to a suitable spectral subspace, as illustrated in Example~\ref{ex:example-3}. In the present construction, we formulate the necessary regularity condition directly by requiring the pointwise spinor scalar products to define bounded sesquilinear forms on the chosen solution subspace.}. To represent these forms by bounded linear operators on a common Hilbert space, we seek a linear subspace
\[
    \mathcal V\subset(\Gamma(\mathbb S \M)\cap \H^m)
\]
such that, for every $x\in\M$, the restriction of $B_x$ to $\mathcal V$ is bounded with respect to the norm induced by $\H^m$. Explicitly, for every $x\in\M$, we require the existence of a constant $C_x>0$ such that
\[
    |B_x(\psi_i,\psi_j)|
    \leq C_x\|\psi_i\|_{\H^m}\|\psi_j\|_{\H^m}
    \qquad
    \text{for all }\psi_i,\psi_j\in\left(\overline{\mathcal V}^{\H^m}\cap\Gamma(\mathbb S \M)\right).
\]
We then define
\[
    \H:=\overline{\mathcal V}^{\,\H^m}
\]
and denote its scalar product, inherited from $\H^m$, by $\bra\cdot,\cdot\ket_{\H}$.
Since $\mathcal V$ is dense in $\H$ by construction, the boundedness of $B_x$ on $\mathcal V$ ensures that it extends uniquely to a bounded sesquilinear form on $\H$. By the Riesz representation theorem, there exists a unique bounded linear operator $F(x)\in\mathrm L(\H)$ satisfying
\begin{equation} \label{eq:local_correlation_operator}
    \bra\psi_i,F(x)\psi_j\ket_{\H}
    =
    B_x(\psi_i,\psi_j)
    \qquad
    \text{for all }\psi_i,\psi_j\in\H,
\end{equation}
where $B_x$ now denotes its continuous extension. Since $B_x$ is Hermitian, the operator $F(x)$ is self-adjoint. Moreover, since $B_x$ is induced by the spin scalar product on $\mathbb S_x\M$, which has signature $(n,n)$, the operator $F(x)$ has rank at most $\dim \mathbb S_x\M$, with at most $n$ positive and at most $n$ negative eigenvalues. We therefore obtain the local correlation map
\begin{equation}\label{eq:local_correlation_map}
    F:\M\longrightarrow\F_n,
    \qquad x\longmapsto F(x).
\end{equation}
Since the matrix elements of $F(x)$ on the dense subspace $\mathcal{V}$ depend smoothly on $x$, and $\H$ is separable, the local correlation map $F$ is Borel measurable with respect to the operator-norm topology on $\F_n$. Accordingly, we can now use this map to define a measure on $\F_n$. Let $\mu_g$ denote the metric volume measure on $\M$. We define the measure $\rho$ as the pushforward of $\mu_g$ under the local correlation map $F$, i.e., $
\rho:=F_*\mu_g.$
Explicitly, for every Borel set $\Omega\subset\F_n$,
\begin{equation} \label{eq:push-forward_measure_explicit}
    \rho(\Omega)=\mu_g\big(F^{-1}(\Omega)\big).
\end{equation}
This construction gives rise to the \ac{cfs}
\[
    (\H,\F_n,\rho).
\]
This construction of a \ac{cfs} depends on the choice of the Hilbert space $\H\subset\H^m$. For a spin isometry to act on the resulting \ac{cfs}, this choice must be compatible with the induced unitary transformation. Let $\widetilde\Phi$ be a Dirac-compatible spin isometry inducing the unitary operator $U^m$ on $\H^m$. We require that the chosen Hilbert subspace $\H$ be invariant under $U^m$, i.e.,
\[
    U^m\H=\H.
\]
Consequently, the restriction $U:=U^m|_{\H}$ defines a unitary operator on $\H$ and we refer to $\H$ as invariant under $U^m$.

\begin{Lemma}[Equivariance of the Local Correlation Map] \label{lem:local_correlation_equivariance}
    Let $(\M,g)$ be a globally hyperbolic spin spacetime and $\widetilde\Phi$ a Dirac-compatible spin isometry covering a time-orientation preserving isometry $\Phi:\M\to\M$. Let $U^m:\H^m\to\H^m$ be the induced unitary operator. Suppose that $\H$ is an invariant subspace under $U^m$ such that the local correlation map $F:\M\to\F_n$ is well-defined on all of $\M$. Then
    \[
        U^{-1}F(x)U=F(\Phi^{-1}(x))
        \qquad\text{for all }x\in\M,
    \]
    where $U:=U^m|_{\H}$.
\end{Lemma}

\begin{proof}
    Let $\psi_i , \psi_j \in \H\cap\Gamma(\mathbb S \M)$. Since $U : \H \to \H$ is unitary, we have
    \[
        \bra
        \psi_i,U^{-1}F(x)U\psi_j
        \ket_{\H}
        =
        \left\bra
        U\psi_i,F(x)U\psi_j
        \right\ket_{\H}.
    \]
    By Equation \eqref{eq:local_correlation_operator},
    \[
        \bra
        U\psi_i,F(x)U\psi_j
        \ket_{\H}
        =
        \prec (U\psi_i)(x)\mid (U\psi_j)(x)\succ_x.
    \]
    By Definition \ref{def:solution_spin_map} the action of the spin isometry on spinor sections \eqref{eq:spinaction}, 
    \[
        (U\psi)(x)
        =
        \widetilde{\Phi}_{\Phi^{-1}(x)}
        \big(
        \psi(\Phi^{-1}(x))
        \big).
    \]
    we obtain
    \begin{align*}
        \prec (U\psi_i)(x)\mid (U\psi_j)(x)\succ_x
        &=
        \prec
        \widetilde{\Phi}|_{\Phi^{-1}(x)}\big(\psi_i(\Phi^{-1}(x))\big)
        \mid
        \widetilde{\Phi}|_{\Phi^{-1}(x)}\big(\psi_j(\Phi^{-1}(x))\big)
        \succ_{x}
        \\
        &=
        \prec\psi_i(\Phi^{-1}(x))\mid\psi_j(\Phi^{-1}(x))\succ_{\Phi^{-1}(x)}.
    \end{align*}
    where the last equality follows from the preservation of the spin scalar product by the spin isometry. Thus, applying Equation \eqref{eq:local_correlation_operator} once more, we obtain
    \[
        \bra\psi_i,U^{-1}F(x)U\psi_j\ket_{\H}
        =
        \prec\psi_i(\Phi^{-1}(x))\mid\psi_j(\Phi^{-1}(x))\succ_{\Phi^{-1}(x)}
        =
        \bra\psi_i,F(\Phi^{-1}(x))\psi_j\ket_{\H}.
    \]
    Since this holds for all $\psi_i,\psi_j\in\H\cap\Gamma(\mathbb S\M)$, which is dense in $\H$, and both operators are bounded, the identity extends by continuity to all of $\H$. Hence,
    \[
        U^{-1}F(x)U
        =
        F(\Phi^{-1}(x))
    \]
    or, equivalently,
    \[
        UF(x)U^{-1}=F(\Phi(x)).
    \]
\end{proof}

The lemma expresses the central geometric idea of the construction: the action of the spacetime symmetry on points of $\M$ is represented on the \ac{cfs} side by unitary conjugation of the corresponding local correlation operators. Together with the invariance of the metric volume measure under spacetime isometries, this equivariance implies the following result.
\begin{Thm} \label{thm:spacetime_to_cfs_symmetry}
    Let $(\M,g)$ be a globally hyperbolic spin spacetime and $\widetilde{\Phi}$ a Dirac-compatible spin isometry covering a time-orientation-preserving isometry $\Phi:\M\to\M$. Let $U^m:\H^m\to\H^m$ be the induced unitary operator.
    
    Suppose that $\H\subset\H^m$ is an invariant Hilbert subspace under $U^m$, such that the local correlation map $F:\M\to\F_n$ is well-defined on all of $\M$. Let $\rho=F_*\mu_g$ be the associated measure on $\F_n$, where $\mu_g$ denotes the metric volume measure on $\M$.
    
    Then $U:=U^m|_{\H}$ is a symmetry of the \ac{cfs} $(\H,\F_n,\rho)$ in the sense of Definition \ref{def:symm}, i.e.,
    \[
        \rho(U\Omega U^{-1})=\rho(\Omega)
    \]
    for every Borel set $\Omega\subset\F_n$.
\end{Thm}

\begin{proof}
    Let $\Omega\subset\F_n$ be a Borel set. By the definition of the pushforward measure $\rho=F_*\mu_g$, we have
    \[
        \rho(U\Omega U^{-1})
        =
        \mu_g\big(F^{-1}(U\Omega U^{-1})\big).
    \]
    By the equivariance of the local correlation map established in Lemma \ref{lem:local_correlation_equivariance}, we have
    \[
        F^{-1}(U\Omega U^{-1})
        =
        \Phi\big(F^{-1}(\Omega)\big).
    \]
    Substituting this relation into the expression for the pushforward measure, we obtain
    \[
        \rho(U\Omega U^{-1})
        =
        \mu_g\big(\Phi(F^{-1}(\Omega))\big).
    \]
    Since $\Phi$ is an isometry of $(\M,g)$, it preserves the metric volume measure $\mu_g$. In particular, for every measurable set $A\subset\M$,
    \[
        \mu_g(\Phi(A))=\mu_g(A).
    \]
    Applying this to $A=F^{-1}(\Omega)$, we conclude that
    \[
    \begin{aligned}
        \rho(U\Omega U^{-1})
        =\mu_g\big(\Phi(F^{-1}(\Omega))\big)
        =\mu_g\big(F^{-1}(\Omega)\big)
        =\rho(\Omega).
    \end{aligned}
    \]
    Thus, $\rho$ is invariant under conjugation by $U$. Since $U$ is unitary on $\H$, it follows from Definition \ref{def:symm} that $U$ is a symmetry of the \ac{cfs} $(\H,\F_n,\rho)$.
\end{proof}

\begin{Corollary}[Group Action] \label{cor:group_action}
    Let $(\M,g)$ be a globally hyperbolic spin spacetime, and let $(G, *)$ be a group acting on the spinor bundle $\mathbb S\M$ by Dirac-compatible spin isometries $\widetilde{\Phi}_h$, each covering a time-orientation-preserving spacetime isometry $\Phi_h$.

    Suppose that there exists a common Hilbert subspace $\H\subset\H^m$, invariant under the induced unitary operators $U_h^m$ for every $h\in G$, such that the associated local correlation map $F:\M\to\F_n$ is well-defined on all of $\M$.

    Then the restricted operators $U_h:=U_h^m|_{\H}$ define a unitary representation of $G$ on $\H$. Moreover, each $U_h$ is a symmetry of the associated \ac{cfs} $(\H,\F_n,\rho)$, where $\rho=F_*\mu_g$, in the sense of Definition \ref{def:symm}.
\end{Corollary}

\begin{Remark} \label{rem:internal} 
    The preceding construction also accommodates internal symmetries, namely spin isometries $\widetilde{\Phi}:\mathbb S\M\to\mathbb S\M$ covering the identity map on spacetime,
    \[
        \Phi=\mathrm{id}_{\M},
    \]
    while potentially acting nontrivially on the spinor fibres. If such a spin isometry is Dirac-compatible and preserves the chosen Hilbert subspace $\H$, then Lemma \ref{lem:local_correlation_equivariance} gives
    \[
        U^{-1}F(x)U=F(x)
        \qquad\text{for all }x\in\M,
    \]
    where $U=U^m|_{\H}$. Thus, internal symmetries satisfy the stronger condition
    \[
        [U,F(x)]=0
        \qquad\text{for all }x\in\M,
    \]
    and therefore leave each local correlation operator invariant.
    However, the possible internal symmetries are constrained by the spin geometry and the Dirac equation. In the standard irreducible complex Dirac spinor representation, compatibility with Clifford multiplication implies, by Schur's lemma \cite[Proposition 4]{Schur_Lemma}, that any fibrewise transformation covering the identity must be of the form
    \[
        \widetilde{\Phi}_x
        =
        e^{i\theta(x)} \, \mathrm{id}_{\mathbb S_x\M}.
    \]
    Preservation of the spin scalar product restricts the scalar factor to a phase. 
    Dirac compatibility requires the phase to be locally constant. Since \(\M\) is connected, it must therefore be constant on all of \(\M\).
    Consequently the internal spin isometries admitted by the standard irreducible Dirac representation form a global $U(1)$ symmetry group.

    This restriction does not exclude local Lorentz invariance. Local Lorentz transformations correspond to changes of local orthonormal frames and their associated spin frames, leaving the underlying geometric structures unchanged. In contrast, the internal symmetries considered here are automorphisms of the spinor bundle preserving the fixed Clifford multiplication.
    
    This restriction relies on the irreducibility of the Clifford representation. If one considers multiple equivalent Dirac sectors, the commutant of Clifford multiplication becomes larger, allowing non-Abelian internal symmetry groups. For instance, for $N$ identical Dirac sectors, the internal transformations preserving Clifford multiplication and the spin scalar product form $U(N)$. In particular, this allows suitable unitary representations of the Standard Model gauge group $SU(3)\times SU(2)\times U(1)$ to be accommodated within the present construction, provided the corresponding transformations are Dirac-compatible and preserve the chosen Hilbert subspace $\H$.
    
    The distinction between the irreducible and multiple-sector cases is reflected in their action on the \ac{cfs} operator space. In the irreducible case, global phase transformations induce unitary operators $U=e^{i\theta}I_{\H}$, whose conjugation action is trivial on all of $\F_n$, since
    \[
        (e^{i\theta}I_{\H})\,A\,(e^{-i\theta}I_{\H})=A
        \qquad\text{for every }A\in\F_n.
    \]
    In contrast, non-Abelian internal symmetries may act nontrivially by conjugation on the full operator space $\F_n$. Nevertheless, by Lemma \ref{lem:local_correlation_equivariance}, they leave every local correlation operator $F(x)$ invariant and consequently fix the support of the measure $\rho=F_*\mu_g$ pointwise. Hence, in both cases, the induced unitary transformations define symmetries of the associated $\ac{cfs}$, even though their actions on the full operator space may differ.
\end{Remark}

\subsubsection{Mapping of Continuous Spacetime Symmetries} \label{sec:spacetime_symmetry_mapping}

While internal symmetries act trivially on the spacetime manifold, spacetime isometries generally induce nontrivial transformations of the local correlation operators. Of particular interest are continuous isometries generated by Killing vector fields.

\begin{Lemma}[Existence of Dirac-Compatible Spin Lifts]\label{lem:spin_lift}
    Let $(\M,g)$ be a globally hyperbolic Lorentzian spin manifold, and let $K$ be a complete Killing vector field on $\M$ with associated isometric flow $(\Phi_t)_{t\in\mathbb R}$.
    
    Then there exists a one-parameter group of spin isometries $(\widetilde\Phi_t)_{t\in\mathbb R}$ covering $(\Phi_t)_{t\in\mathbb R}$, such that
    \[
        D_{\M}^m\circ\widetilde\Phi_t
        =
        \widetilde\Phi_t\circ D_{\M}^m
        :
        \Gamma(\mathbb S\M)
        \longrightarrow
        \Gamma(\mathbb S\M),
        \qquad \forall t\in\mathbb R.
    \]
    In particular, every $\widetilde\Phi_t$ is Dirac-compatible.
\end{Lemma}

\begin{proof}
    Since $K$ is complete, its flow $(\Phi_t)_{t\in\mathbb R}$ is globally defined. As $K$ is Killing, each $\Phi_t$ is an isometry. Moreover, the flow is connected to the identity and therefore preserves the orientation and time orientation of $\M$.

    As shown in \cite[\text{Section 3}]{Compatible_Lift}, the isometric flow generated by a Killing vector field lifts to a one-parameter group of automorphisms of the spin bundle. These automorphisms preserve the spin scalar product, intertwine Clifford multiplication,  and hence define a one-parameter group of spin isometries $(\widetilde{\Phi}_t)_{t\in\mathbb R}$ covering $(\Phi_t)_{t\in\mathbb R}$. For a systematic treatment of this construction, see also \cite{Godina_and_Matteucci}; a related construction of local group actions from complete Killing fields is given in \cite[Proposition~5.3]{SigSymm}.

    Since each $\Phi_t$ is an isometry, it preserves the Levi-Civita connection. Consequently, the corresponding lift $\widetilde\Phi_t$ preserves the induced spin connection. Thus, for every $X\in\Gamma(T\M)$, $\psi\in\Gamma(\mathbb S\M)$, and $x\in\M$,
    \[
        \nabla^{\mathbb S\M}_{d\Phi_t|_x(X_x)}
        \bigl(\widetilde\Phi_t\psi\bigr)\big|_{\Phi_t(x)}
        =
        \widetilde\Phi_{t,x}
        \left(
        \nabla^{\mathbb S\M}_{X_x}\psi\big|_x
        \right).
    \]    
    Let $(e_0,\ldots,e_{d-1})$ be a local orthonormal frame around $x\in\M$, and let $(e'_0,\ldots,e'_{d-1})$ be its pushforward under $\Phi_t$. Since $\Phi_t$ is an isometry, $(e_0',\ldots,e_{d-1}')$ is an orthonormal frame around $\Phi_t(x)$. Using the preservation of the spin connection and the intertwining property of Clifford multiplication, Definition \ref{def:Dirac_op} gives
    \[
    \begin{alignedat}{2}
        D_{\M}(\widetilde\Phi_t\psi)\big|_{\Phi_t(x)}
        &=
        i\sum_{a=0}^{d-1}\varepsilon_a\,
        \gamma_{\M}(e_a')
        \nabla^{\mathbb S\M}_{e_a'}
        (\widetilde\Phi_t\psi)\big|_{\Phi_t(x)}
        &&=
        i\sum_{a=0}^{d-1}\varepsilon_a\,
        \gamma_{\M}(e_a')
        \widetilde\Phi_{t,x}
        \left(\nabla^{\mathbb S\M}_{e_a}\psi\big|_x\right)
        \\
        &=
        i\sum_{a=0}^{d-1}\varepsilon_a\,
        \widetilde\Phi_{t,x}
        \left(
        \gamma_{\M}(e_a)
        \nabla^{\mathbb S\M}_{e_a}\psi\big|_x
        \right)
        &&=
        \widetilde\Phi_{t,x}
        \left(D_{\M}\psi\big|_x\right),
    \end{alignedat}
    \]
    Since the mass term is a scalar multiple of the identity, the same identity holds for the massive Dirac operator:
    \[
        D_{\M}^m(\widetilde\Phi_t\psi)\big|_{\Phi_t(x)}
        =
        \widetilde\Phi_{t,x}
        \left(D_{\M}^m\psi\big|_x\right).
    \]
    Thus,
    \[
        D_{\M}^m\circ\widetilde\Phi_t
        =
        \widetilde\Phi_t\circ D_{\M}^m
        :
        \Gamma(\mathbb S\M)
        \longrightarrow
        \Gamma(\mathbb S\M).
    \]
    As $\Phi_t$ is an isometry, it maps compact subsets of Cauchy surfaces to compact subsets of Cauchy surfaces. Therefore, $\widetilde\Phi_t$ maps spatially compact solutions of the massive Dirac equation bijectively onto themselves. Hence every $\widetilde\Phi_t$ is Dirac-compatible.
\end{proof}
To eventually apply Stone's theorem, we need the strong continuity of the induced unitary group (see also \cite[Theorem~3.8]{SigSymm}).
\begin{Prp} \label{prp:stong_continuity}
    Let $(\M,g)$ be a globally hyperbolic Lorentzian spin manifold, and let $K$ be a complete Killing vector field with associated isometric flow $(\Phi_t)_{t\in\mathbb R}$. Let $(\widetilde\Phi_t)_{t\in\mathbb R}$ be the one-parameter group of Dirac-compatible spin isometries provided by Lemma $\ref{lem:spin_lift}$, and denote by $(U_t^m)_{t\in\mathbb R}$ the corresponding one-parameter unitary group on $\H^m$. Then $(U_t^m)_{t\in\mathbb R}$ is strongly continuous. Moreover, for every spacelike Cauchy hypersurface $\Sigma$, the induced unitary group $(U_{\Sigma,t}^m)_{t\in\mathbb R}$ on $\H_\Sigma$ is also strongly continuous.
\end{Prp}
    
\begin{proof}
    To establish strong continuity, we need to show that
    \[
    \lim_{t\to0}\|U_t^m\psi-\psi\|_{\H^m}=0
    \]
    for every \(\psi\in\H^m\).
    
    We first prove this statement for all \(\psi\in\Gamma_{\mathrm{sc}}^m(\mathbb S\M)\). Recall that 
    $\left.U_t^m\right|_{\Gamma_{\mathrm{sc}}^m(\mathbb S\M)}
    =
    \left.\widetilde\Phi_t\right|_{\Gamma_{\mathrm{sc}}^m(\mathbb S\M)}
    $ where the action of \(\widetilde\Phi_t\) on spinor sections is given by \eqref{eq:spinaction} as
    \[
    (\widetilde\Phi_t\psi)(x)
    =
    \left.\widetilde\Phi_t\right|_{\Phi_{-t}(x)}
    \bigl(\psi(\Phi_{-t}(x))\bigr).
    \]
    Since the flow and its spin lift depend smoothly on \(t\) and reduce to the identity at \(t=0\), we obtain
    \[
    \lim_{t\to0}(\widetilde\Phi_t\psi)(x)=\psi(x)
    \]
    pointwise for every \(x\in\M\). Moreover, smooth dependence implies uniform convergence on compact subsets. To make use of this fact we fix a spacelike Cauchy surface \(\Sigma\) and let
    \[
    K_0:=\operatorname{supp}(R_\Sigma^m\psi),
    \]
    which is compact. By finite propagation of the massive Dirac equation \cite[Theorem 4.5.2]{CFS_Book},  we have
    \[
    \operatorname{supp}\psi\subset J(K_0)
    \]
    and since \(\Phi_t\) preserves the causal structure, also 
    \[
    \operatorname{supp}(U_t^m\psi)
    \subset J(\Phi_t(K_0)).
    \]
    For any \(\varepsilon>0\), the set
    \[
    C:=\bigcup_{|t|\leq\varepsilon}\Phi_t(K_0)
    \]
    is compact by continuity of the flow. Global hyperbolicity then implies that
    \[
    L:=J(C)\cap\Sigma \quad \text{ is compact with }\quad \operatorname{supp}(R_\Sigma^mU_t^m\psi)\subset L
    \qquad (|t|\leq\varepsilon).
    \]
    Since the supports of \(R_\Sigma^m U_t^m\psi\) and \(R_\Sigma^m\psi\) are contained in \(L\) for \(|t|\leq\varepsilon\), the definition of the scalar product \eqref{eq:solution_scalar_product} gives
    \[
    \begin{aligned}
    \lim_{t\to0}&\|U_t^m\psi-\psi\|_{\H^m}^2\\
    &=
    \lim_{t\to0}\int_L
    \prec (R_\Sigma^m U_t^m\psi-R_\Sigma^m \psi)(x)
    \mid
    \gamma_{\M}(\nu_\Sigma(x))
    (R_\Sigma^m U_t^m\psi-R_\Sigma^m \psi)(x)\succ_x
    \,d\mu_\Sigma(x)\\
    &=0.
    \end{aligned}
    \]
    The first equality follows from the restriction of the support to $L$ while the second follows from the uniform convergence on \(L\), the smoothness of the fibrewise scalar product, and the compactness of \(L\). This completes the argument on $\Gamma_{\mathrm{sc}}^m(\mathbb S\M)$. Now let \(\psi\in\H^m\) be arbitrary. For any \(\widetilde\psi\in\Gamma_{\mathrm{sc}}^m(\mathbb S\M)\), the triangle inequality and unitarity of \(U_t^m\) give
    \[
    \begin{aligned}
    \|U_t^m\psi-\psi\|_{\H^m}
    &\leq
    \|U_t^m(\psi-\widetilde\psi)\|_{\H^m}
    +\|U_t^m\widetilde\psi-\widetilde\psi\|_{\H^m}
    +\|\widetilde\psi-\psi\|_{\H^m}\\
    &=
    2\|\psi-\widetilde\psi\|_{\H^m}
    +\|U_t^m\widetilde\psi-\widetilde\psi\|_{\H^m}.
    \end{aligned}
    \]
    Taking the limit as \(t\to0\), we obtain
    \[
    \limsup_{t\to0}
    \|U_t^m\psi-\psi\|_{\H^m}
    \leq
    2\|\psi-\widetilde\psi\|_{\H^m}.
    \]
    Since \(\Gamma_{\mathrm{sc}}^m(\mathbb S\M)\) is dense in \(\H^m\) by Definition \ref{def:Sol_Hilbert_space}, the right-hand side can be made arbitrarily small. As the norm is non-negative, it follows that
    \[
    0\leq
    \liminf_{t\to0}\|U_t^m\psi-\psi\|_{\H^m}
    \leq
    \limsup_{t\to0}\|U_t^m\psi-\psi\|_{\H^m}
    =0.
    \]
    Thus, the limit exists and
    \[
    \lim_{t\to0}\|U_t^m\psi-\psi\|_{\H^m}=0
    \qquad
    \text{for every }\psi\in\H^m.
    \]
    The group property then implies strong continuity at every \(t\in\mathbb R\).
    Finally, by Definition \ref{def:hypersurface_unitary},
    \[
    U_{\Sigma,t}^m
    =
    R_\Sigma^m U_t^m(R_\Sigma^m)^{-1}.
    \]
    Since \(R_\Sigma^m\) is unitary by Proposition \ref{prop:restriction_unitary}, the induced group \((U_{\Sigma,t}^m)_{t\in\mathbb R}\) is also strongly continuous.
\end{proof}

This then leads us to the following corollary to Theorem \ref{thm:spacetime_to_cfs_symmetry}.

\begin{Corollary} \label{cor:spacetime_symmetry}
    Let $(\M,g)$ be a globally hyperbolic Lorentzian spin manifold, and let $K$ be a complete Killing vector field with associated isometric flow $(\Phi_t)_{t\in\mathbb R}$.
    Let $(\widetilde\Phi_t)_{t\in\mathbb R}$ be the one-parameter group of Dirac-compatible spin isometries provided by Lemma \ref{lem:spin_lift}, and denote by $U_t^m$ the corresponding unitary operators on $\H^m$.
    Suppose there exists a common invariant Hilbert subspace $\H\subset\H^m$ such that
    \[
        U_t^m\H=\H
        \qquad\text{for all }t\in\mathbb R,
    \]
    and the local correlation map $F:\M\to\F_n$ is well-defined on all of $\M$.
  Then the restrictions \(U_t:=\left.U_t^m\right|_{\H}\) form a strongly continuous one-parameter unitary group on \(\H\). Moreover, each \(U_t\) is a symmetry of the associated \ac{cfs} \((\H,\F_n,\rho)\), where \(\rho=F_*\mu_g\).
\end{Corollary}

\begin{proof}
    By Lemma \ref{lem:spin_lift}, the spin isometries $\widetilde\Phi_t$ are Dirac-compatible and cover the time-orientation-preserving isometries $\Phi_t$. Proposition \ref{prop:solution_spin_isometry_unitary} therefore gives a one-parameter group of unitary operators $U_t^m$ on $\H^m$.

    By assumption, $\H$ is invariant under every $U_t^m$, so the restrictions $U_t$ form a one-parameter unitary group on $\H$. Moreover, by Proposition \ref{prp:stong_continuity}, for every \(\psi\in\H\),
\[
\lim_{t\to0}\|U_t\psi-\psi\|_{\H}
=
\lim_{t\to0}\|U_t^m\psi-\psi\|_{\H^m}
=0.
\]
Thus, the restricted group is strongly continuous.

    Finally, Theorem \ref{thm:spacetime_to_cfs_symmetry} implies that each $U_t$ preserves the measure $\rho$ under conjugation. Hence, \((U_t)_{t\in\mathbb R}\) is a strongly continuous one-parameter group of symmetries of the associated \ac{cfs}.
\end{proof}

While the spin lift constructed in Lemma  \ref{lem:spin_lift} is not unique (cf. the discussion of lift ambiguities in \cite[Section~5.1]{SigSymm}), we now show that the induced action on the \ac{cfs} is independent of this choice.

\begin{Prp}[Uniqueness of the Induced \ac{cfs} Action]\label{prp:uniqueness_of_action}
   Let \((\M,g)\) be a connected, globally hyperbolic Lorentzian spin manifold, and let \(K\) be a complete Killing vector field with associated isometric flow \((\Phi_t)_{t\in\mathbb R}\). Let $(\widetilde\Phi_t)_{t\in\mathbb R}$ and $(\widehat\Phi_t)_{t\in\mathbb R}$ be two one-parameter groups of Dirac-compatible spin isometries covering the same flow, and denote their induced unitary operators on $\H^m$ by $U_t^m$ and $\widehat U_t^m$, respectively.
   
    Suppose there exists a Hilbert subspace $\H\subset\H^m$ satisfying
    \[
        U_t^m\H=\H
        \qquad\text{for all }t\in\mathbb R,
    \]
    such that the local correlation map $F:\M\to\F_n$ is well-defined on all of $\M$.
    Then $\H$ is also invariant under every $\widehat U_t^m$, and the restrictions
    \[
        U_t:=U_t^m|_{\H},
        \qquad
        \widehat U_t:=\widehat U_t^m|_{\H}
    \]
    induce the same \ac{cfs} symmetry, i.e.,
    \[
        U_t F(x)U_t^{-1}
        =
        \widehat U_t F(x)\widehat U_t^{-1}
        \qquad
        \text{for all }x\in\M,\quad t\in\mathbb R.
    \]
\end{Prp}

\begin{proof}
    For each $t\in\mathbb R$, the composition $\widehat\Phi_t^{-1}\circ\widetilde\Phi_t$ is a Dirac-compatible spin isometry covering the identity on $\M$. By Remark \ref{rem:internal}, there exists a constant phase $e^{i\omega t}\in U(1)$ such that
    \[
        \widetilde\Phi_t=e^{i\omega t}\widehat\Phi_t.
    \]
    Consequently, the induced unitary operators satisfy $U_t^m=e^{i\omega t}\widehat U_t^m$. Since $\H$ is invariant under $U_t^m$, it follows that
    \[
        \widehat U_t^m\H
        =e^{-i\omega t}U_t^m\H
        =\H.
    \]
    Thus both one-parameter groups restrict to unitary groups on the same Hilbert subspace $\H$, with $U_t=e^{i\omega t}\widehat U_t$. The scalar phases cancel under conjugation, giving
    \[
        U_t F(x)U_t^{-1}
        =
        \widehat U_t F(x)\widehat U_t^{-1}
    \]
    for every $x\in\M$ and $t\in\mathbb R$.
\end{proof}

We have thus shown that, under suitable assumptions, \ac{cfs} constructed from globally hyperbolic spin spacetimes admitting a one-parameter group of isometries possess the kind of one-parameter unitary symmetry studied in Section \ref{sec:observable_&_evolution}.

\subsubsection{Hamiltonian Description and Finite-Dimensional Invariant Subspaces} \label{sec:finite_dim_invariant}

By Proposition \ref{prp:stong_continuity}, the one-parameter unitary group \((U_t^m)_{t\in\mathbb R}\) is strongly continuous. Hence, by Stone's Theorem \ref{thm:Stone}, there exists a self-adjoint generator $H^m$ on $\H^m$ such that
\[
    U_t^m=e^{itH^m},\qquad \forall t\in\mathbb R.
\]
For a Hilbert subspace $\H\subset\H^m$ to be invariant under the one-parameter group, it must be preserved by the action of $H^m$. Equivalently, defining the orthogonal projection $P_{\H}$ onto $\H$, we require
\[
    [P_{\H},H^m]=0.
\]
Since the self-adjoint generator $H^m$ may be unbounded, its domain $\mathcal D(H^m)$ is in general only a dense subspace of $\H^m$. The commutation condition must therefore be understood as
\[
    P_{\H}\mathcal D(H^m)\subseteq\mathcal D(H^m),
    \qquad
    H^mP_{\H}\psi=P_{\H}H^m\psi
    \quad\text{for all }\psi\in\mathcal D(H^m).
\]
This is equivalent to the invariance of $\H$ under $U_t^m$ for all $t\in\mathbb R$.
In particular, the existence of a nontrivial finite-dimensional invariant subspace requires $H^m$ to possess a nonempty point spectrum. Indeed, on such a subspace the restriction $H=H^m|_{\H}$ is an everywhere-defined self-adjoint operator, and the spectral theorem guarantees an orthonormal basis of eigenvectors $\psi_i$, satisfying
\[
    H\psi_i=\lambda_i\psi_i,
    \qquad
    U_t\psi_i=e^{it\lambda_i}\psi_i.
\]
Conversely, any finite collection of eigenvectors of $H^m$ spans a finite-dimensional invariant subspace. However, the construction of the local correlation map additionally requires these eigenvectors to admit smooth representatives, which is not guaranteed in general. 

\subsubsection{Hamiltonian Description on Static Spacetimes}\label{sec:staticspacetimes}

In the case of static spacetimes this question can be addressed explicitly. The static geometry provides a preferred time coordinate and allows the spacetime Dirac operator to be decomposed into a geometric time derivative and a spatial differential operator. The Dirac equation thus becomes a relation between the generator of the geometric symmetry and the generator of the \ac{cfs} symmetry. 

Let $(\M,g)$ be a connected, globally hyperbolic static Lorentzian spin manifold, with
\[
    \M=\mathbb R\times\Sigma,
    \qquad
    g=\beta^2\,dt^2-h,
\]
where $\beta>0$ and the Riemannian metric $h$ are independent of $t$. The hypersurfaces $\Sigma_t:=\{t\}\times\Sigma$ are spacelike Cauchy hypersurfaces. The vector field
\[
    K=\partial_t
\]
is a complete timelike Killing vector field, with flow $(\Phi_t)_{t\in\mathbb R}$. We write
\[
    \nu=\beta^{-1}K
\]
for the future-directed unit normal to the hypersurfaces and identify the reference Cauchy hypersurface with $\Sigma=\Sigma_0$.

As established in Lemma \ref{lem:spin_lift}, the Killing flow admits a one-parameter group of Dirac-compatible spin-isometric lifts. We distinguish the canonical spin lift \((\widetilde\Phi_t^{\mathrm{can}})_{t\in\mathbb R}\), constructed in \cite[Section 3]{Compatible_Lift} by lifting the induced action on the orthonormal frame bundle to the principal spin bundle, starting at the identity. The corresponding spinorial Lie derivative is defined by
\begin{equation}\label{eq:spinorialLie}
   (\mathcal L_K^{\mathrm{spin}}\psi)(x)
=
\left.\frac{d}{dt}\right|_{t=0}
\widetilde\Phi_{-t,\Phi_t(x)}^{\mathrm{can}}
\psi(\Phi_t(x)). 
\end{equation}
We get the following decomposition of the Dirac operator.
\begin{Prp}\label{prp:dirac_decomposition}
    Let \((\M,g)\) be a connected, globally hyperbolic static Lorentzian spin manifold and let \(m\geq 0\).
On \(\Gamma(\mathbb S\M)\), the massive Dirac operator admits the decomposition
\begin{equation}\label{eq:Dirac_decomposition}
    D_{\M}^m = \frac{1}{\beta}\gamma_{\M}(\nu) \left(i\mathcal L_K^{\mathrm{spin}}-H_D^m\right), 
\end{equation} 
where the spatial Dirac Hamiltonian is given by
\begin{equation} \label{eq:Spatial_Hamiltonian}
    H_D^m = \frac12\{\beta,H_\Sigma\} + m\beta\gamma_{\M}(\nu).
\end{equation}
Here \(\{\cdot,\cdot\}\) denotes the anticommutator and \(H_\Sigma\) the hypersurface Dirac operator.
\end{Prp}

\begin{proof}
    Choose a local $h$-orthonormal frame $e_1,\ldots,e_{d-1}$ on $\Sigma$ and extend it along the Killing flow. Together with $\nu=\beta^{-1}K$, this gives a local pseudo-orthonormal frame for the spacetime metric. 
Since the metric is static, the hypersurfaces \(\Sigma_t\) are totally geodesic. In particular,
\[
\nabla_X^{\M}\nu=0
\qquad\text{for all }X\in T\Sigma_t.
\]
Hence, the induced spin connection on the restricted spinor bundle agrees with the spacetime spin connection in tangential directions:
\[
\nabla_X^\Sigma(\psi|_{\Sigma_t})
=
\left.(\nabla_X^{\mathbb S\M}\psi)\right|_{\Sigma_t}.
\]
In the normal direction, the spinorial Lie derivative formula, together with the acceleration of the unit normal, gives
\[
\begin{aligned}
\nabla_\nu^{\mathbb S\M}
&=\frac{1}{\beta}\mathcal L_K^{\mathrm{spin}}
-\frac12\gamma_{\M}(\nu)\gamma_{\M}
\left(\nabla_\nu^{\M}\nu\right)\\
&=\frac{1}{\beta}\mathcal L_K^{\mathrm{spin}}
+\frac12\alpha\left(\operatorname{grad}_h(\log\beta)\right),
\end{aligned}
\]
where we introduced the spatial Clifford multiplication
\[
\alpha(X):=-\gamma_{\M}(\nu)\gamma_{\M}(X),
\qquad X\in T\Sigma_t.
\]
Starting with Definition~\ref{def:Dirac_op}, we then get
\begin{align*}
         D_{\M}^{m}
        &=
        i\sum_{\mu=0}^{d-1}
        \varepsilon_{\mu}\,\gamma_{\M}(e_{\mu})\nabla^{\mathbb S\M}_{e_{\mu}}-m\,\mathbb{I}\\
        &=i\gamma_{\M}(\nu)\nabla^{\mathbb S\M}_\nu
-i\sum_{a=1}^{d-1}\gamma_{\M}(e_a)\nabla^\Sigma_{e_a}
-m\,\mathbb{I}
\\
&=\frac{i}{\beta}\gamma_{\M}(\nu)\mathcal L_K^{\mathrm{spin}}
+\frac{i}{2}\gamma_{\M}(\nu)
\alpha\big(\operatorname{grad}_h\log\beta\big)
-\gamma_{\M}(\nu)H_\Sigma-m\,\mathbb{I}
\\
&=\frac1\beta\gamma_{\M}(\nu)
\left(
i\mathcal L_K^{\mathrm{spin}}
-\beta H_\Sigma
+\frac{i}{2}\alpha(\operatorname{grad}_h\beta)
-m\beta\gamma_{\M}(\nu)
\right)
\\
&=\frac1\beta\gamma_{\M}(\nu)
\left(
i\mathcal L_K^{\mathrm{spin}}
-\frac12\{\beta,H_\Sigma\}
-m\beta\gamma_{\M}(\nu)
\right)\,.
\end{align*}
In the third line we introduced the hypersurface Dirac operator
\[
H_\Sigma:=-i\sum_{a=1}^{d-1}\alpha(e_a)\nabla^\Sigma_{e_a}
\]
and in the last line we used the Leibniz rule
\[
H_\Sigma(\beta\psi)
=\beta H_\Sigma\psi-i\alpha(\operatorname{grad}_h\beta)\psi\qquad \,,
\]
which holds both for \(\psi\in\Gamma(\mathbb S\M)\), with \(H_\Sigma\) acting tangentially, and for \(\psi\in\Gamma(\mathbb S\M|_{\Sigma_t})\). Thus, defining \(H_D^m\) as in \eqref{eq:Spatial_Hamiltonian}, the final expression gives precisely the decomposition \eqref{eq:Dirac_decomposition}.
\end{proof}

For smooth spatially compact solutions \(\psi\in\Gamma_{\mathrm{sc}}^m(\mathbb S\M)\), Proposition \ref{prp:dirac_decomposition} gives
\begin{equation}\label{eq:symmetry_relation}
  D_{\M}^m\psi=0
\quad\Longleftrightarrow\quad
i\mathcal L_K^{\mathrm{spin}}\psi=H_D^m\psi.  
\end{equation}
Thus, the Dirac equation directly relates the infinitesimal generator of the geometric symmetry to the spatial Dirac Hamiltonian. The corresponding Hamiltonian identification and essential self-adjointness of the infinitesimal symmetry generator were established in \cite[Section~6.2 and Theorem~3.9(i)]{SigSymm}, the latter using Chernoff's theorem. Here, we obtain essential self-adjointness by combining Stone's theorem with an invariant-core argument.
\begin{Thm}[Connecting Killing Vector Fields to \ac{cfs} Symmetry Generators]\label{thm:generator_correspondence}
    Let \((\M,g)\) be a connected, globally hyperbolic static Lorentzian spin manifold with complete timelike Killing vector field \(K=\partial_t\), and let \(m\geq 0\).

    Let \(H_{\mathrm{can}}^m\) denote the self-adjoint Stone generator on \(\H^m\), introduced in Section \ref{sec:finite_dim_invariant}, corresponding to the canonical spin lift of the Killing flow. Then
\[
H_{\mathrm{can}}^m
=
\overline{\left.H_D^m\right|_{\Gamma_{\mathrm{sc}}^m(\mathbb S\M)}}.
\]
Moreover, the Stone generator \(H_\Sigma^{m,\mathrm{can}}\) of the induced strongly continuous unitary group on \(\H_\Sigma\) introduced in Definition \ref{def:hypersurface_unitary} satisfies
\[
H_\Sigma^{m,\mathrm{can}}
=
R_\Sigma^m H_{\mathrm{can}}^m(R_\Sigma^m)^{-1}
=
\overline{\left.H_D^m\right|_{\Gamma_c(\mathbb S\M|_\Sigma)}}.
\]
Here, the overlines denote the operator closures in \(\H^m\) and \(\H_\Sigma\), respectively.

For any other one-parameter group of Dirac-compatible spin lifts covering the same Killing flow, with corresponding Stone generator \(H^m\), there exists \(\omega\in\mathbb R\) such that
\[
H^m=H_{\mathrm{can}}^m+\omega I,
\qquad
R_\Sigma^m H^m(R_\Sigma^m)^{-1}
=
H_\Sigma^{m,\mathrm{can}}+\omega I.
\]
All equalities hold as identities of self-adjoint operators, including their domains.   
\end{Thm}

\begin{proof}
    Let \(\psi\in\Gamma_{\mathrm{sc}}^m(\mathbb S\M)\). By the definitions of the canonical unitary action in Proposition \ref{prop:solution_spin_isometry_unitary} and the spinorial Lie derivative \eqref{eq:spinorialLie}, we have, for every \(x\in\M\),
\[
\left.\frac{d}{dt}\right|_{t=0}
\big(U_t^{m,\mathrm{can}}\psi\big)(x)
=
-\big(\mathcal L_K^{\mathrm{spin}}\psi\big)(x).
\]
We first show that, for every \(\psi\in\Gamma_{\mathrm{sc}}^m(\mathbb S\M)\), this pointwise derivative also exists in the norm of \(\H^m\).
By Lemma \ref{lem:spin_lift}, the canonical spin lift commutes with \(D_{\M}^m\). Differentiating this relation at \(t=0\), and using that \(\mathcal L_K^{\mathrm{spin}}\) is a local differential operator, we obtain
\[
\mathcal L_K^{\mathrm{spin}}\psi
\in\Gamma_{\mathrm{sc}}^m(\mathbb S\M).
\]
Now define
\[
q_t:=
\frac{U_t^{m,\mathrm{can}}\psi-\psi}{t}
+\mathcal L_K^{\mathrm{spin}}\psi.
\]
As in the proof of Proposition \ref{prp:stong_continuity}, there exists a compact set \(L\subset\Sigma\) containing the supports of \(R_\Sigma^m U_t^{m,\mathrm{can}}\psi\) for all sufficiently small \(|t|\). By the pointwise derivative established above, \(L\) also contains the support of \(R_\Sigma^m\mathcal L_K^{\mathrm{spin}}\psi\). Hence
\[
\operatorname{supp}(R_\Sigma^m q_t)\subseteq L.
\]
On each fibre \(\mathbb S_x\M\), \(x\in\Sigma\), the positive-definite scalar product induces the norm
\[
|v|_x^2
:=
\prec v\mid\gamma_{\M}(\nu)v\succ_x,
\qquad v\in \mathbb S_x\M.
\]
Smooth dependence on \(t\) implies that the difference quotients converge uniformly on \(L\), so
\[
\lim_{t\to0}\sup_{x\in L}|(R_\Sigma^m q_t)(x)|_x
=0\,.
\]
Using the definition of the scalar product \eqref{eq:solution_scalar_product}, we therefore obtain
\[
\begin{aligned}
\lim_{t\to0}
\left\|
\frac{U_t^{m,\mathrm{can}}\psi-\psi}{t}
+\mathcal L_K^{\mathrm{spin}}\psi
\right\|^2_{\H^m}&=\lim_{t\to0}\|q_t\|_{\H^m}^2\\
&=
\lim_{t\to0}\int_L
\prec (R_\Sigma^m q_t)(x)
\mid\gamma_{\M}(\nu)(R_\Sigma^m q_t)(x)
\succ_x\,d\mu_\Sigma(x)\\
&\leq\lim_{t\to0}
\mu_\Sigma(L)
\sup_{x\in L}|(R_\Sigma^m q_t)(x)|_x^2=0\,.
\end{aligned}
\]
Thus, the pointwise derivative also exists in the \(\H^m\)-norm. Hence the smooth spatially compact sections on $\mathbb{S}\M$ are contained in the domain of  the Stone derivative
\[
\Gamma_{\mathrm{sc}}^m(\mathbb S\M)
\subseteq \mathcal D(H_{\mathrm{can}}^m),
\]
and for every \(\psi\in\Gamma_{\mathrm{sc}}^m(\mathbb S\M)\), we obtain the identification 
\[
iH_{\mathrm{can}}^m\psi
=\left.\frac{d}{dt}\right|_{t=0}U_t^{m,\mathrm{can}}\psi
=-\mathcal L_K^{\mathrm{spin}}\psi
=iH_D^m\psi\,,
\]
Here, the first equality follows from the characterization of the generator in Stone's theorem \ref{thm:Stone}, while the last follows from the special form of the Dirac equation in static spacetimes \eqref{eq:symmetry_relation}.

By Definition \ref{def:Sol_Hilbert_space}, \(\Gamma_{\mathrm{sc}}^m(\mathbb S\M)\) is dense in \(\H^m\). Moreover, Lemma \ref{lem:spin_lift} implies that this space is invariant under \(U_t^{m,\mathrm{can}}\) for every \(t\in\mathbb R\), and we have just established that
\[
\Gamma_{\mathrm{sc}}^m(\mathbb S\M)
\subseteq\mathcal D(H_{\mathrm{can}}^m).
\]
Thus, by the invariant-core theorem for strongly continuous semigroups \cite[Proposition~II.1.7]{Engel_Nagel}, applied to the restriction of the unitary group to \(t\geq0\), \(\Gamma_{\mathrm{sc}}^m(\mathbb S\M)\) is a core for \(H_{\mathrm{can}}^m\). Since \(H_D^m\) agrees with \(H_{\mathrm{can}}^m\) on this core, we conclude that
\[
H_{\mathrm{can}}^m
=
\overline{\left.H_D^m\right|_{\Gamma_{\mathrm{sc}}^m(\mathbb S\M)}}.
\]
This establishes the first assertion of the theorem, including equality of the operator domains.

By Definition \ref{def:hypersurface_unitary}, the canonical unitary group on \(\H_\Sigma\) is given by
\[
U_{\Sigma,t}^{m,\mathrm{can}}
=R_\Sigma^m U_t^{m,\mathrm{can}}(R_\Sigma^m)^{-1}.
\]
Since \(R_\Sigma^m\) is unitary by Proposition \ref{prop:restriction_unitary}, uniqueness of the Stone generator implies
\[
H_\Sigma^{m,\mathrm{can}}
=R_\Sigma^m H_{\mathrm{can}}^m(R_\Sigma^m)^{-1}.
\]
Moreover, the well-posedness of the Dirac Cauchy problem gives
\[
R_\Sigma^m\big(\Gamma_{\mathrm{sc}}^m(\mathbb S\M)\big)
=\Gamma_c(\mathbb S\M|_\Sigma).
\]
Since \(H_D^m\), as defined in \eqref{eq:Spatial_Hamiltonian}, acts tangentially to the Cauchy hypersurfaces, its action commutes with restriction to \(\Sigma\). Consequently, for every \(\psi\in\Gamma_{\mathrm{sc}}^m(\mathbb S\M)\),
\[
R_\Sigma^m H_{\mathrm{can}}^m\psi
=R_\Sigma^m H_D^m\psi
=H_D^m R_\Sigma^m\psi.
\]
Since \(\Gamma_{\mathrm{sc}}^m(\mathbb S\M)\) is a core for \(H_{\mathrm{can}}^m\), its image under the unitary map \(R_\Sigma^m\) is a core for \(H_\Sigma^{m,\mathrm{can}}\). We therefore conclude that
\[
H_\Sigma^{m,\mathrm{can}}
=R_\Sigma^m H_{\mathrm{can}}^m(R_\Sigma^m)^{-1}
=\overline{\left.H_D^m\right|_{\Gamma_c(\mathbb S\M|_\Sigma)}}.
\]
This establishes the second assertion of the theorem, including equality of the operator domains.

Finally, let \((U_t^m)_{t\in\mathbb R}\) be the unitary group induced by any other one-parameter group of Dirac-compatible spin lifts covering the same Killing flow. By Remark \ref{rem:internal}, the two groups differ by a scalar phase \(e^{i\omega t}\) for some \(\omega\in\mathbb R\). Hence,
\[
U_t^m=e^{i\omega t}U_t^{m,\mathrm{can}}
=e^{it(H_{\mathrm{can}}^m+\omega I)}.
\]
By uniqueness of the Stone generator, we obtain
\[
H^m=H_{\mathrm{can}}^m+\omega I,
\qquad
\mathcal D(H^m)=\mathcal D(H_{\mathrm{can}}^m).
\]
Consequently, using Proposition \ref{prop:restriction_unitary} and the preceding identification on \(\H_\Sigma\),
\[
R_\Sigma^m H^m(R_\Sigma^m)^{-1}
=H_\Sigma^{m,\mathrm{can}}+\omega I.
\]
Since the identity operator is bounded, the addition of \(\omega I\) does not change the operator domain. This establishes the final assertion of the theorem.

\end{proof}

To interpret this result as a dynamical equation on the fixed hypersurface Hilbert space \(\H_\Sigma\), we identify the spinor bundles over the hypersurfaces \(\Sigma_t\) using the canonical Killing flow. More precisely, a local spinor frame \(s_0\) on \(\Sigma\) is extended by
\[
s_t(\Phi_t(x))
=
\widetilde\Phi_{t,x}^{\mathrm{can}}(s_0(x)).
\]
In these transported frames, the spinorial Lie derivative is represented on spinor components by the ordinary time derivative,
\[
\mathcal L_K^{\mathrm{spin}}=\partial_t.
\]
Consequently, the Dirac equation takes the form
\begin{equation}\label{eq:SchrödingerForm}
   i\partial_t\psi=H_D^m\psi\,.
\end{equation}
Let \(\psi_0\in\Gamma_c(\mathbb S\M|_\Sigma)\) be initial data. Using the geometric flow map to identify \(\H_{\Sigma_t}\) with the fixed Hilbert space \(\H_\Sigma\), we can pull back the Dirac evolution to \(\H_\Sigma\) by defining
\[
\psi_t
:=\G_{\Sigma,\Sigma_t}E_{\Sigma_t,\Sigma}^m\psi_0.
\]
By Definition \ref{def:hypersurface_unitary} and the group property, we have
\[
\G_{\Sigma,\Sigma_t}E_{\Sigma_t,\Sigma}^m
=
\left(U_{\Sigma,t}^{m,\mathrm{can}}\right)^{-1}
=
U_{\Sigma,-t}^{m,\mathrm{can}}.
\]
Consequently, Theorem \ref{thm:generator_correspondence} gives
\[
\psi_t
=
U_{\Sigma,-t}^{m,\mathrm{can}}\psi_0
=
e^{-itH_\Sigma^{m,\mathrm{can}}}\psi_0.
\]
Thus, the Dirac equation takes the form of a Schrödinger evolution equation on the fixed Hilbert space \(\H_\Sigma\),
\[
i\frac{d}{dt}\psi_t
=
H_\Sigma^{m,\mathrm{can}}\psi_t,
\qquad
\psi_{t=0}=\psi_0.
\]
By Theorem \ref{thm:generator_correspondence}, \(H_\Sigma^{m,\mathrm{can}}\) is self-adjoint. Consequently, the evolution is well-defined and strongly continuous for arbitrary initial data in \(\H_\Sigma\), while for initial data in \(\mathcal D(H_\Sigma^{m,\mathrm{can}})\), it satisfies the Schrödinger equation in the Hilbert-space norm.

Furthermore, Theorem \ref{thm:generator_correspondence} also clarifies how the freedom in choosing the spin lift is reflected in this Hamiltonian description. Changing the spin lift shifts the corresponding self-adjoint generator by a constant multiple of the identity,
\[
H_\Sigma^{m,\mathrm{can}}
\longmapsto H_\Sigma^{m,\mathrm{can}}+\omega I,
\qquad \omega\in\mathbb R.
\]
Consequently, the Schrödinger evolution changes only by an overall phase,
\[
e^{-it(H_\Sigma^{m,\mathrm{can}}+\omega I)}
=
e^{-i\omega t}e^{-itH_\Sigma^{m,\mathrm{can}}},
\]
while the spectrum undergoes a uniform shift,
\[
\sigma(H_\Sigma^{m,\mathrm{can}}+\omega I)
=
\sigma(H_\Sigma^{m,\mathrm{can}})+\omega.
\]
To understand the (in)significance of this spectral shift for the induced \ac{cfs} symmetry, consider a finite-dimensional invariant Hilbert subspace \(\H\subset\H^m\) satisfying the assumptions of Corollary \ref{cor:spacetime_symmetry}. Let \(H\) denote the generator of the restricted unitary group \(U_t=e^{itH}\), and let \(\{\psi_j\}\) be an orthonormal eigenbasis of \(H\), with \(H\psi_j=\lambda_j\psi_j\). The induced action on the local correlation operators then satisfies
\[
\left\langle \psi_i,
e^{itH}F(x)e^{-itH}\psi_j
\right\rangle_{\H}
=
e^{it(\lambda_i-\lambda_j)}
\left\langle\psi_i,F(x)\psi_j\right\rangle_{\H}.
\]
Thus, the action on the local correlation operators depends only on the differences between the eigenvalues of the generator. Under a change of spin lift, all eigenvalues are shifted by the same constant, leaving these differences unchanged,
\[
(\lambda_i+\omega)-(\lambda_j+\omega)
=\lambda_i-\lambda_j.
\]
Therefore, the independence of the induced \ac{cfs} symmetry from the choice of spin lift, established in Proposition \ref{prp:uniqueness_of_action}, is reflected in the Hamiltonian description by the fact that only relative spectral values enter the action on the local correlation operators. Thus, only relative energies enter the induced action on the local correlation operators, while the choice of the zero point of energy is irrelevant. Although this is expected from the usual Hamiltonian formulation of quantum mechanics, it provides a further consistency check for the interpretation of \ac{cfs} as a purely relational theory, as developed in \cite{PC}.

The same spectral representation has a further consequence for the evolution of the local correlation operators. By Lemma \ref{lem:local_correlation_equivariance}, we have
\[
F(\Phi_t(x))=e^{itH}F(x)e^{-itH}.
\]
Since \(\H\) is finite-dimensional, the matrix elements of \(F(\Phi_t(x))\) are finite sums of oscillatory terms with frequencies given by differences of the eigenvalues of \(H\). Consequently, for every \(x\in\M\), the map
\[
t\longmapsto F(\Phi_t(x))
\]
is almost periodic in operator norm. In particular, for every \(\varepsilon>0\), there exist arbitrarily large times \(t\) such that
\[
\|F(\Phi_t(x))-F(x)\|<\varepsilon.
\]
Thus, the local correlation operators exhibit recurrence along the Killing flow, even if the underlying spacetime flow is not periodic. An even stronger conclusion holds if the spectral differences are commensurate, meaning that there exists \(\Omega>0\) such that
\[
\lambda_i-\lambda_j\in\Omega\mathbb Z
\qquad\text{for all }i,j.
\]
In this case, the evolution of the local correlation operators is exactly periodic with period \(T=2\pi/\Omega\).
\begin{Remark}
For finite-dimensional invariant Hilbert subspaces, the local correlation operators of any compact spatial region remain within a compact set under the Killing flow. Consequently, the pushforward of the full spacetime volume measure is not locally finite. While this does not affect Lemma \ref{lem:local_correlation_equivariance} or the symmetry of the resulting \ac{cfs} in Theorem \ref{thm:spacetime_to_cfs_symmetry}, the measure does not satisfy the local-finiteness requirement of the standard infinite-volume formulation of the causal action principle. Moreover, the $\sigma$-finiteness required for the evolution results of Section \ref{sec:observable_&_evolution} is not guaranteed.  If the local correlation map is exactly periodic, restricting the spacetime volume measure to one period yields a finite \ac{cfs} with the induced one-parameter group of symmetries. In infinite-dimensional invariant Hilbert subspaces, the compactness argument no longer applies.
\end{Remark}
\subsubsection{Examples}

Having established the framework for relating spacetime symmetries to \ac{cfs} symmetries, we now illustrate the construction with three examples. The first two demonstrate how suitable finite-dimensional invariant Hilbert subspaces can be constructed, the third demonstrates that the construction can also apply to infinite-dimensional Hilbert spaces.

\begin{example}[Stationary Spacetimes with Compact Cauchy Surface] \label{ex:example-1}

Let \((\M,g)\) be a globally hyperbolic stationary spacetime with compact Cauchy surfaces. In \cite[Theorem~1.1]{Islam2023Gutzwiller}, it was shown that the infinitesimal generator of the Killing-induced unitary evolution on the Hilbert space of Dirac solutions has purely discrete real spectrum with finite multiplicities. Moreover, elliptic regularity implies that the corresponding eigenvectors of the self-adjoint generator are smooth.

Consequently, the span \(\H\) of any finite collection of eigenvectors of the self-adjoint generator is a finite-dimensional invariant subspace of \(\H^m\). Since these solutions are smooth and \(\H\) is finite-dimensional, the pointwise spin scalar products define bounded sesquilinear forms on \(\H\) at every \(x\in\M\). Hence, the local correlation map \(F:\M\to\F_n\) is well-defined, and Corollary \ref{cor:spacetime_symmetry} implies that the induced unitary evolution defines a one-parameter group of \ac{cfs} symmetries.
\end{example}

\begin{example}[Minkowski Spacetime in a Periodic Box] \label{ex:example-2}

In \cite{FinsterIsidro2023} Finster and Isidro consider a \ac{cfs} from a spatially periodic Minkowski spacetime. Following this construction, we consider the free massive Dirac equation on
\[
\mathcal{M}_L = \mathbb{R} \times \mathbb{T}_L^3,
\quad
\mathbb{T}_L^3 := \frac{\mathbb{R}^3}{(2L\mathbb{Z})^3},
\quad \text{ with metric }\quad
g = dt^2 - \sum_{j=1}^3 (dx^j)^2.
\]
The Dirac Hamiltonian given in \eqref{eq:Spatial_Hamiltonian} simplifies in this setting to
\[
H_D^m
=
-i\sum_{j=1}^{3}\alpha(\partial_{x^j})\nabla^\Sigma_{\partial_{x^j}}
+m\gamma_{\M}(\partial_t).
\]
The periodic boundary conditions on the spinors corresponding to the identification of the opposite faces, restrict the allowed spatial momenta to
\[
k\in\mathcal L_L:=\frac{\pi}{L}\mathbb Z^3,
\]
giving the energy eigenvalues
\[
\lambda_\pm(k)=\pm\sqrt{|k|^2+m^2}.
\]
For a parameter \(0<\varepsilon<m^{-1}\), consider the subspace spanned by the negative-energy modes satisfying
\[
\lambda_-(k)\geq-\varepsilon^{-1}
\qquad\Longleftrightarrow\qquad
|k|\leq\sqrt{\varepsilon^{-2}-m^2}.
\]
Since only finitely many lattice momenta satisfy this bound, the corresponding modes span a finite-dimensional invariant subspace consisting of smooth Dirac solutions.

By Theorem \ref{thm:generator_correspondence}, the Dirac Hamiltonian is unitarily equivalent to the self-adjoint generator of the canonical Killing-induced unitary evolution. Hence, the finite-dimensional subspace $\H$ spanned by the selected modes is invariant under this evolution. Since $\H$ consists of smooth Dirac solutions and is finite-dimensional, the pointwise spin scalar products define bounded sesquilinear forms on $\H$. Consequently, the local correlation map is well-defined, and Corollary \ref{cor:spacetime_symmetry} implies that the induced unitary evolution defines a strongly continuous one-parameter group of symmetries of the associated \ac{cfs}.

As discussed at the end of Section \ref{sec:staticspacetimes}, the finite dimensionality of $\H$ implies that the evolution of the local correlation operators along the Killing flow is almost periodic. In their study of cosmological effects from \ac{cfs}, Finster and Isidro \cite{FinsterIsidro2023} go a step further by imposing exact periodicity in time with a period $T$, interpreted as the lifetime of the universe. Since the energy eigenvalues are generally incommensurate, such periodicity cannot be obtained from the unmodified Dirac sea. To overcome this obstruction, they introduce a smearing of the mass shell, allowing the energy spectrum to be compatible with temporal periodicity. The resulting modification of the Dirac sea gives rise to corrections which they interpret as contributions to dark matter and dark energy.
\end{example}

\begin{example}[Minkowski Spacetime] \label{ex:example-3}

We now drop the spatial periodicity of Example \ref{ex:example-2} and consider the free massive Dirac equation on Minkowski spacetime
\[
\M=\mathbb R\times\mathbb R^3,
\qquad
g=dt^2-\sum_{j=1}^3(dx^j)^2.
\]
The Dirac Hamiltonian is unchanged, but the allowed spatial momenta now form a continuum. Under the spatial Fourier transform, the Hamiltonian acts by multiplication with
\[
\widehat H_D^m(k)
=
\sum_{j=1}^3 \alpha(\partial_{x^j})\,k_j
+m\gamma_{\M}(\partial_t),
\qquad k\in\mathbb R^3,
\]
with energy eigenvalues
\[
\lambda_\pm(k)=\pm\sqrt{|k|^2+m^2}.
\]
The construction of causal fermion systems from regularized Dirac seas in Minkowski spacetime has been studied extensively; see, e.g., \cite{Continuum_Limit_CFS} and references therein. As in Example \ref{ex:example-2}, we consider the particularly simple case of a sharp ultraviolet cut-off and consider the negative-energy spectral subspace
\[
\H_\varepsilon
=
\operatorname{Ran}
\mathbf 1_{[-\varepsilon^{-1},-m]}(H_D^m).
\]
The corresponding momentum support is restricted to
\[
|k|\leq\sqrt{\varepsilon^{-2}-m^2}.
\]
Unlike the periodic case, however, this cutoff does not yield a finite-dimensional subspace, since the allowed momenta form a continuum. 

By Theorem \ref{thm:generator_correspondence}, the generator of the canonical Killing-induced unitary evolution is unitarily equivalent to the Dirac Hamiltonian. Since both the spectral projection
\[
P_\varepsilon
=\mathbf{1}_{[-\varepsilon^{-1},-m]}(H_D^m)
\]
and the unitary evolution \(e^{itH_D^m}\) are functions of the same self-adjoint operator, the spectral theorem implies that they commute,
\[
P_\varepsilon e^{itH_D^m}
=e^{itH_D^m}P_\varepsilon.
\]
Consequently, the range of \(P_\varepsilon\) is invariant under the Killing-induced unitary evolution.

To verify that the local correlation map is well-defined, we first note that in the standard Minkowski spinor representation, for any \(\psi_i,\psi_j\in\H_\varepsilon\) and \(x\in\M\), the spin scalar product satisfies
\[
\left|\prec\psi_i(x)\mid\psi_j(x)\succ_x\right|
\leq
\|\psi_i(x)\|_{\mathbb C^4}\,
\|\psi_j(x)\|_{\mathbb C^4}.
\]
To bound the right-hand side, we denote the momentum ball determined by the ultraviolet cutoff by
\[
B_\varepsilon
:=
\left\{
k\in\mathbb R^3:
|k|\leq\sqrt{\varepsilon^{-2}-m^2}
\right\}.
\]
For any \(\psi\in\H_\varepsilon\), Fourier inversion gives
\[
\psi(t,x)
=
\frac{1}{(2\pi)^{3/2}}
\int_{B_\varepsilon}
e^{ik\cdot x}
e^{-it\widehat H_D^m(k)}
\widehat\psi(0,k)\,d^3k.
\]

Since \(e^{-it\widehat H_D^m(k)}\) is unitary, the Cauchy–Schwarz inequality yields
\[
\begin{aligned}
\|\psi(t,x)\|_{\mathbb C^4}
&\leq
\frac{1}{(2\pi)^{3/2}}
\int_{B_\varepsilon}
\|\widehat\psi(0,k)\|_{\mathbb C^4}\,d^3k
\\
&\leq
\left(
\frac{\operatorname{vol}(B_\varepsilon)}{(2\pi)^3}
\right)^{1/2}
\|\psi\|_{\H^m},
\end{aligned}
\]
where the last step uses Plancherel's theorem and the unitary identification of the Dirac Hilbert space with the hypersurface Hilbert space.

Combining these estimates, we obtain
\[
\left|
\prec\psi_i(x)\mid\psi_j(x)\succ_x
\right|
\leq
\frac{\operatorname{vol}(B_\varepsilon)}{(2\pi)^3}
\|\psi_i\|_{\H^m}
\|\psi_j\|_{\H^m}.
\]
Thus, the pointwise spin scalar products define bounded sesquilinear forms on \(\H_\varepsilon\), uniformly in \(x\in\M\). Moreover, the bounded momentum support ensures that the corresponding Dirac solutions are smooth. Hence, the local correlation map is well-defined on all of \(\M\), and Corollary \ref{cor:spacetime_symmetry} implies that the Killing-induced unitary evolution defines a strongly continuous one-parameter group of symmetries of the associated \ac{cfs}.

Unlike the preceding examples, the restricted Dirac Hamiltonian has purely absolutely continuous spectrum, and the evolution of the nonzero local correlation operators is therefore not almost periodic in operator norm.
\end{example}

\section{Conclusion and Outlook} \label{sec:conclusion_&_outlook}

In this work, we investigated the interplay between observables and symmetries in \ac{cfs}. We established that position observables transform covariantly under unitary symmetries of a \ac{cfs}, satisfying
\[
\mathcal O(UAU^{-1})=U\mathcal O(A)U^{-1}.
\]
To give this abstract covariance relation a geometric interpretation, we investigated the correspondence between spacetime isometries and symmetries of causal fermion systems constructed from solutions of the Dirac equation. We showed that, under suitable assumptions on the chosen Hilbert subspace and the local correlation map, Dirac-compatible spin isometries induce unitary transformations that preserve the measure of the associated \ac{cfs}. In particular, the local correlation map intertwines the action of spacetime isometries with unitary conjugation of the local correlation operators. This provides a concrete answer, for the \ac{cfs} considered here, to the question raised at the end of \cite[Section 2.4]{QRF_CG} concerning the extent to which symmetries of effective physical models induce corresponding symmetries of their associated correlation geometries.

Combining these results establishes a direct correspondence between the geometric action of spacetime symmetries and the covariance of position observables in the associated \ac{cfs}. Complete Killing flows give rise to strongly continuous one-parameter groups of \ac{cfs} symmetries, whose generators determine the Heisenberg-type evolution of the position observables. In static spacetimes, the Dirac equation provides the crucial link between the infinitesimal generator of the geometric symmetry, expressed through the spinorial Lie derivative along the Killing vector field, and the generator of the induced unitary \ac{cfs} symmetry. In particular, the Dirac equation yields
\[
i\mathcal L_K^{\mathrm{spin}}\psi=H_D^m\psi,
\]
allowing us to identify the generator of the induced unitary symmetry with the spatial Dirac Hamiltonian. Thus, the geometric symmetry of spacetime, the Dirac dynamics, the unitary evolution on the Hilbert space, and the transformation of localization observables are connected within a common framework.

The correspondence established in this work relies on spin isometries preserving both Clifford multiplication and the spin scalar product. As discussed in Remark \ref{rem:internal}, on the connected spacetimes considered here, these requirements restrict internal spin isometries to global $U(1)$ phase transformations in the irreducible Dirac representation, while larger internal symmetry groups become possible when multiple equivalent Dirac sectors are considered. However, the requirement of preserving Clifford multiplication may exclude other physically relevant symmetries. In even spacetime dimensions, for example, the massless Dirac equation admits global chiral transformations generated by the chirality operator. These transformations preserve the space of massless Dirac solutions and act unitarily on the corresponding solution Hilbert space, but they neither commute with Clifford multiplication nor generally preserve the indefinite spin scalar product. Consequently, they are not spin isometries in the sense of Definition \ref{def:spin_isometry}, and our symmetry-mapping theorem does not apply. This raises the question of whether transformations that preserve the solution Hilbert space, but not the underlying spin geometry in the required sense, can nevertheless induce symmetries of the associated \ac{cfs}. 

An important conceptual distinction, made explicit by our symmetry-mapping results, concerns the different roles of local Lorentz invariance and internal symmetries. As already emphasized in \cite{QRF_CG} the construction relies on intrinsic geometric structures and is therefore independent of the choice of coordinates and local frames. In particular, local Lorentz transformations merely change the description of the underlying spin geometry and do not appear as additional unitary symmetries of the resulting CFS. 
Internal symmetries, by contrast, can act nontrivially on the Hilbert space of Dirac solutions, while leaving the local correlation operators invariant. This distinction is particularly interesting in view of the noncompactness of the Lorentz group: local Lorentz invariance is already incorporated into the geometric construction, without requiring the local Lorentz group to be represented as a group of physical unitary symmetries on the CFS Hilbert space. Investigating the implications of this separation for quantization, as well as its extension to local internal gauge symmetries, provides an interesting direction for future research.

The extension to multiple Dirac sectors also provides a natural starting point for investigating gauge interactions. Indeed, the emergence of Standard Model gauge fields from the causal action principle is already established in the continuum limit of \ac{cfs}  \cite{CFS_Book}, while the structure of the underlying internal symmetries has also been explored in connection with octonionic descriptions of particle physics \cite{Octonian}. A natural direction for future work is therefore to extend the symmetry correspondence established here to the correlation geometry of interacting Dirac systems and investigate how to construct explicit unitary transformations that realize gauge transformations as unitary equivalences between the associated \ac{cfs} \cite[Definition 2.13]{QRF_CG}. This also provides a setting for studying averaging procedures over the corresponding unitary identifications. Another interesting possibility, recently suggested by Marcianò \cite{Antonino}, is to investigate whether connections of Wilczek--Zee type \cite{Wilczek_Zee}  can be constructed directly from the correlation structures of a \ac{cfs}, providing a complementary geometric perspective on gauge symmetries.

The construction of correlation geometries from Lorentzian spacetimes raises a fundamental question concerning the distinguished role of the Dirac equation in \ac{cfs}. Is the Dirac equation singled out by the requirements of this construction, or could other field equations provide an equally suitable starting point? The construction requires a Hilbert space on which correlations can be defined throughout spacetime while depending on as few choices as possible. The careful distinction between the global solution Hilbert space, the hypersurface Hilbert spaces, and their unitary identifications in Section \ref{sec:symmetry_mapping} highlights why the Dirac equation is particularly well suited to this construction. Its positive scalar product is independent of the chosen Cauchy hypersurface, ensuring that the probabilistic interpretation of local experiments does not depend on the hypersurface used to describe the state. Moreover, this structure allows us to identify unitarily the solution spaces of different Dirac equations on the same spacetime, including equations with different gauge potentials. While these identifications generally depend on the chosen hypersurface, gauge transformations provide intrinsic unitary equivalences between the corresponding correlation geometries. This raises the question of whether the existence of such a hypersurface-independent Hilbert-space structure, together with suitable geometric assumptions, distinguishes Dirac-type equations from other possible dynamics.

In \cite{FinsterIsidro2023}, Finster and Isidro found that imposing exact periodicity on the \ac{cfs} arising from Minkowski spacetime in a box leads to cosmological-constant-like contributions to the Euler–Lagrange equations. Unlike global de Sitter spacetime, its static patch does not admit spatially compact Cauchy hypersurfaces and is not covered by the examples considered in the present paper. In light of \cite{Continuum_FF}, investigating the interacting domain of de Sitter spacetime is therefore a natural direction for future work. What makes this particularly interesting is that the thermodynamics of the cosmological horizon allows for the derivation of a scaling relation between the Planck length (identified with the UV regularization length) and the cosmological constant (setting an IR scale) under the assumption that every occupied state carries a fixed spatial volume. 

This assumption is closely related to the equipartition conjecture formulated in \cite{PC}, which proposes that minimizing the causal action leads to an equal distribution of spacetime volume among one-particle states. Building on earlier work on the support of minimizing \ac{cfs} \cite[Lemma 3.4]{Felix_Patrick}, we establish in Theorem \ref{thm:exhaustion} that for a minimizer of the causal action in a finite-dimensional \ac{cfs},  every normalized one-particle state carries strictly positive spacetime volume. Establishing an appropriate formulation of equipartition for spacetimes with finite spatial volume but infinite temporal extent could provide a connection to the thermodynamic argument above. Moreover, equipartition would imply that the total position observable is proportional to the identity, allowing the positive, additive position observables constructed here to be normalized into a POVM by a constant rescaling. Together with the covariance established in the present paper, this suggests investigating whether these position observables can serve as \ac{qrf} observables in the construction of symmetry-invariant relational observables, as developed in \cite{Leon_Relative_Time,Leon_QRF} thereby strengthening the connection between correlation geometry and \acp{qrf} first explored in \cite{QRF_CG}.

\appendix
\section{Positivity of the Total Position Observable} \label{app:positivity}

We begin with a standard positivity result for the causal action, which will be used in the proof of the subsequent theorem.
\begin{Lemma}\label{lem:positive_action}
    Let \((\H,\F_n,\rho)\) be a \ac{cfs} with \(\dim\H<\infty\), and suppose that \(\rho\) is an admissible measure satisfying the constraints of the causal action principle. Then the causal action is strictly positive, i.e.,
    \begin{equation}
        \S(\rho) >0.
    \end{equation}
\end{Lemma}

\begin{proof}
    From trace constraint,
    \[
        \int_{\F_{n}}\operatorname{tr}(x)d\rho(x)
        \neq
        0,
    \]
    there exists $x_0\in\operatorname{supp}\rho$ such that $\operatorname{tr}(x_0)\neq0$. For $x=y=x_0$, the causal Lagrangian is given as,
    \begin{align}
        \mathcal{L}(x_0,x_0)
        =
        \frac{1}{4n}
        \sum_{i,j=1}^{2n}
        \left(
        |\lambda_i|^2-|\lambda_j|^2
        \right)^2
        =
        \operatorname{tr}(x_0^4)
        -\frac{1}{2n}\operatorname{tr}(x_0^2)^2.
    \end{align}
    By \cite[Definition 6.2.1]{CFS_Book}, $\mathcal{L}(x_0,x_0)\geq0$. Moreover, equality can hold only if $|\lambda_1|=\cdots=|\lambda_{2n}|$. In particular, if $\operatorname{tr}(x_0)\neq0$, the diagonal element of the Lagrangian is strictly positive,
    \[
        \mathcal{L}(x_0,x_0)
        >
        0.
    \]
    
    The causal Lagrangian is continuous \cite[Section 6.2]{CFS_Book}. Therefore, there exists an open neighborhood $\mathcal{U}$ of $x_0$ in $\F_{n}$ such that
    \[
        \mathcal{L}(x,y)>0
        \qquad
        \text{for all }
        x,y\in \mathcal{U}.
    \]
    Since $x_0\in\operatorname{supp}\rho$, every open neighborhood of $x_0$ has positive $\rho$-measure $\rho(\mathcal{U})>0$. Thus, using the non-negativity of the causal Lagrangian, we obtain
    \begin{align}
        \S(\rho)
        =
        \int_{\F_{n}} \int_{\F_{n}}
        \mathcal{L}(x,y) d\rho(x) d \rho(y)
        \geq
        \int_{\mathcal{U}} \int_{\mathcal{U}}
        \mathcal{L}(x,y)d\rho(x)d\rho(y)
        >0.
    \end{align}
    Hence, $\mathcal{S}(\rho)>0$.
\end{proof}
Using this positivity result together with the Euler–Lagrange equations for minimizers of the causal action, we obtain the following property of the total position observable.

\begin{Thm}\label{thm:exhaustion}
    Let \((\H,\F_n,\rho)\) be a \ac{cfs} with \(\dim\H<\infty\), where \(\rho\) is a minimizer of the causal action principle. Let \(\mathcal O(M)\) be the position observable associated with the entire spacetime \(M=\operatorname{supp}\rho\), as defined in Definition \ref{def:observables}. Then,
    \begin{equation}
        \bra u,\O(M)u \ket_{\H}>0 
        \quad
        \forall u\in \H \quad \text{ with } ||u||_{\H}=1 \, .
    \end{equation}
\end{Thm}

\begin{proof}
We consider the Euler–Lagrange function for the constrained causal action principle, as given in \cite[Proposition 2.13]{Continuum_FF},
\begin{equation}\label{eq:constraintEL}
    \ell(x)
=
\int_{\mathcal F_n}
\left(
\mathcal L(x,y)+\kappa |xy|^2
\right)d\rho(y)
-\mathfrak r\,\operatorname{tr}(x)
-\mathfrak s,
\end{equation}
where \(\mathfrak r\) and \(\mathfrak s\) are the Lagrange multipliers associated with the trace and volume constraints, respectively, and \(\kappa\geq 0\) is the multiplier associated with the boundedness constraint.
The causal action \eqref{eq:causal_action} can be expressed in terms of \(\ell\) as
\[
\mathcal S(\rho)=\int_{\mathcal F_n}\ell(x)\,d\rho(x)-\kappa\mathcal T(\rho)+\mathfrak r\,\operatorname{Tr}(\rho)+\mathfrak s\,\rho(\mathcal F_n).
\]
By \cite[Proposition 2.13]{Continuum_FF}, the Euler–Lagrange equations imply
\[
\ell|_M=\inf_{\mathcal F_n}\ell=0.
\]
Integrating over \(\mathcal F_n\) and using the volume and trace constraints, we obtain
\[
0=\int_{\mathcal F_n}\ell(x)\,d\rho(x)=\mathcal S(\rho)+\kappa\mathcal T(\rho)-\mathfrak r-\mathfrak s.
\]
Consequently, by Lemma \ref{lem:positive_action} and \(\kappa\geq 0\),
\begin{equation}
    \mathfrak r+\mathfrak s
=\mathcal S(\rho)+\kappa\mathcal T(\rho)>0.
\end{equation}
 Assume by contradiction that there exists a normalized vector $u\in\mathcal H$ such that $\langle u,O(M)u\rangle_{\mathcal H}=0$. Then,
    \begin{align}
        \int_{M} \bra u,\pi_{x} u \ket \,d\rho(x) 
        =
        0
        &\iff
        \int_{M} ||\pi_{x} u||^{2} \,d\rho(x)
        =
        0\\
        &\implies
        ||\pi_{x}u||^{2}=0 
        \quad
        \text{for $\rho$-almost every $x\in M$}\label{eq:pi_vanish}
    \end{align}
    For any rank-one projection $\F_{n} \ni P_{u}:=|u \ket \bra u|$ and $x\in \F_{n}$, the causal Lagrangian is given by \cite[Lemma 2.2]{Preserver_Paper},
    \begin{equation}
        \L(P_{u},x)=\left( 1-\frac{1}{2n} \right)|\bra u,xu \ket|^{2}.
    \end{equation}
    Moreover, \(P_ux\) has at most one nonzero eigenvalue, equal to \(\langle u,xu\rangle\). Therefore,
\[
|P_ux|^2=|\langle u,xu\rangle|^2,
\]
and the combined Lagrangian becomes
\begin{equation}\label{eq:lagrangian}
\mathcal L(P_u,x)+\kappa|P_ux|^2
=
\left(1-\frac{1}{2n}+\kappa\right)
|\langle u,xu\rangle|^2.
\end{equation}
Since \(\operatorname{tr}(P_u)=1\), the Euler–Lagrange function evaluated at \(P_u\) is
\begin{equation}
    \ell(P_u)=\left(1-\frac{1}{2n}+\kappa\right)\int_{\mathcal F_n}|\langle u,x u\rangle|^2\,d\rho(x)-\mathfrak r-\mathfrak s.
\end{equation}
 and, $x=\pi_{x}x\pi_{x}$, implies,
 \begin{equation}\label{eq:lpu}
     \ell(P_u)=\left(1-\frac{1}{2n}+\kappa\right)\int_{\mathcal F_n}|\langle\pi_xu,x\pi_xu\rangle|^2\,d\rho(x)-\mathfrak r-\mathfrak s.
 \end{equation}
    As, $||\pi_{x}u||^{2}=0$ for $\rho$-almost every $x\in M$, then by Cauchy-Schwarz inequality,
    \begin{equation}
        0
        \leq
        |\bra \pi_{x}u,x\,\pi_{x}u \ket|^{2}
        \leq
        ||x||^{2}||\pi_{x}u||^{4}
    \end{equation}
    Taking integral both side gives,
    \begin{equation}\label{eq:greater_than_zero}
        0
        \leq
        \int_{M} |\bra \pi_{x}u,x\,\pi_{x}u \ket|^{2}\,d\rho(x)
        \leq
        \int_{M} ||x||^{2}||\pi_{x}u||^{4}\,d\rho(x).
    \end{equation}
    Consider compact subsets $M_{i}$ such that $M_{i}\subset M_{i+1}$ with $\bigcup_{i=1 } ^{\infty}M_{i} =M$ then monotonicity of the measure implies $\rho(M_i)\leq \rho(M_{i+1})$ and continuity from below $\lim_{i\rightarrow \infty}\rho(M_i)=\rho(M)$.  As $M_{i}$ are compact subsets and $x\rightarrow||x||$ is continuous, there exists $C_{i}$, such that,
       \[
        C_{i}
        :=
        \max_{x\in M_{i}} ||x||^{2}.
    \]
    As a result we get that for all $i$,
 \begin{equation}\label{eq:less_than_zero}
        \int_{M_{i}}||x||^{2} ||\pi_{x}u||^{4} \, d\rho(x)
        \leq
        C_{i}\int_{M_{i}}||\pi_{x}u||^{4}\,d\rho(x) \stackrel{\eqref{eq:pi_vanish}}{=}0.
   \end{equation}
  In the last step we used that $||\pi_{x}u||$ is a nonnegative measurable function and hence $||\pi_{x}u||^2=0$ $\rho$-almost everywhere implies that $||\pi_{x}u||^4=0$ $\rho$-almost everywhere. Combining Equation \eqref{eq:greater_than_zero} and \eqref{eq:less_than_zero} we get 
\begin{align}
     0&\leq \int_M |\bra \pi_xu,x\pi_xu\ket|^2\,d\rho(x)\nonumber\\ &\leq \int_M \|x\|^2\|\pi_xu\|^4\,d\rho(x) \stackrel{\text{MCT}}{=} \lim_{i\to\infty} \int_{M_i}\|x\|^2\|\pi_xu\|^4\,d\rho(x)=0\, , 
\end{align}
 where the MCT equality is implied by the monotone convergence of $\mathbf 1_{M_i}(x)\|x\|^2\|\pi_xu\|^4 \uparrow \|x\|^2\|\pi_xu\|^4$.  Using this in \eqref{eq:lpu} implies that  $\ell(P_{u}) = -\mathfrak r-\mathfrak s$ and hence $\ell(P_{u})<0=\inf_{\F_{n}}\ell$ which is the desired contradiction.
\end{proof}

\vspace{2em}
\printbibliography

\end{document}